\documentclass[preprints,instruments,review,moreauthors,accept,pdftex]{Definitions/mdpi} 

\firstpage{1} 
\pubvolume{xx}
\issuenum{1}
\articlenumber{5}
\pubyear{2019}
\copyrightyear{2019}
\history{Received: date; Accepted: date; Published: date}

\Title{Wavelength shifters for applications in liquid argon detectors}

\Author{Marcin Ku\'zniak $^{1,*}$\orcidA{} and Andrzej M. Szelc $^{2}$\orcidB{}}

\AuthorNames{Marcin Ku\'zniak and Andrzej M. Szelc}

\address{%
$^{1}$ \quad AstroCeNT, Nicolaus Copernicus Astronomical Center of the Polish Academy of Sciences, ul.~Rektorska 4, 00-614~Warsaw, Poland\\
$^{2}$ \quad School of Physics and Astronomy, University of Edinburgh, Edinburgh EH8 9YL, United Kingdom \\ }
\corres{Correspondence: mkuzniak@camk.edu.pl; Tel.: +48-22-120-1823}

\abstract{Wavelength shifters and their applications for liquid argon detectors have been a subject of extensive R\&D over the past decade.
This work reviews the most recent results in this field. We compare the optical properties and usage details together with the associated challenges for various wavelength shifting solutions. We discuss the current status and potential future R\&D directions for the main classes of wavelength shifters.}

\keyword{Wavelength shifters; Fluorescence; Polyethylene naphthalate; PEN; Tetraphenyl butadiene; TPB; Liquid argon detectors; Noble liquid detectors; VUV detection}

\PACS{29.40.Mc \and 33.50.Dq \and 95.35.+d \and 14.60.Pq}

\begin{document}


\section{Introduction}
Liquid argon (LAr) is used as a scintillator in a number of fundamental physics experiments. These include direct dark matter searches (WArP~\cite{warp}, DEAP~\cite{deap}, MiniCLEAN~\cite{miniclean}, ArDM~\cite{ardm}, DarkSide~\cite{ds20k}), prototype and full-scale neutrino oscillation experiments (ICARUS~\cite{icarus}, MicroBooNE~\cite{microboone}, LArIAT~\cite{lariat}, SBND~\cite{Szelc_2016}, ProtoDUNE~\cite{protodune}, DUNE~\cite{dune}), experiments searching for neutrinoless double beta decay (0$\nu\beta\beta$-decay) (GERDA~\cite{gerdaupgrade}, LEGEND~\cite{legend}) and measuring the coherent elastic neutrino-nucleon scattering (CENNS-10~\cite{Tayloe_2018}). Other applications of LAr-based detectors include medical physics (PET)~\cite{3dpi} and nuclear or homeland security~\cite{Nikkel_2012, alarm}. For more details on detectors based on liquefied noble gasses see~\cite{Aprile} or one of the more recent review papers~\cite{Chepel_2013, marchionni}.

The peak wavelength of the LAr scintillation light is 128~nm, i.e. in vacuum ultraviolet (VUV), where the efficiencies of currently available VUV-optimized photosensors are typically fairly low, reaching at most 15\%~\cite{vuv4} (for the most recent review of the VUV photon detector technology see also~\cite{vuvpd}). Therefore, an efficient wavelength shifter (WLS), a material which absorbs the VUV light and isotropically re-emits it at a longer wavelength, is needed to enable its detection with standard blue-sensitive photosensors: conventional photomultiplier tubes (PMT), silicon photomultipliers (SiPM) or avalanche photodiodes (APD). 
Most wavelength shifting compounds considered typically have very high re-emission efficiencies and fluorescence lifetimes comparable with LAr scintillation timing, leading to an overall increase in light collection in LAr detectors.

With the next generation very large detectors and novel application ideas on the horizon, the past decade has seen extensive R\&D in the field of wavelength shifters for LAr applications. One of the main directions of these efforts is to develop WLS applicable to detectors ranging between hundreds of tons, for the Argo dark matter search~\cite{argo}, and tens of kilotons for the DUNE long-baseline neutrino experiment. These detectors will have surface areas of the inner detector walls ranging between hundreds and thousands of square meters, that may need to be covered with an efficient and stable WLS. 

This scaling-up is non-trivial for the WLS dominantly used in the field until now, e.g. the organic 1,1,4,4-tetraphenyl-1,3-butadiene (TPB). The vacuum deposition method commonly used to deposit TPB films is very difficult to scale because of the associated high vacuum requirements, costs and labor. In consequence, a number of alternative deposition techniques and WLS have been proposed and investigated, among them polyethylene naphthalate (PEN) polymeric films. As an independent effort, WLS tailored to particular background mitigation purposes or optimized for PET or 0$\nu\beta\beta$-decay applications have been studied and developed.

This paper provides an overview of WLS currently used or considered for LAr applications, as well as covers practical aspects of deploying them in the detectors, e.g. in a form of thin coatings. It also highlights the remaining open questions about TPB properties (including its absolute conversion efficiency), which are still debated in the literature, as well as collecting and comparing data on its alternatives. Detailed description of the energy transfer mechanisms for each of the materials discussed is beyond the scope of this paper; it is, nevertheless, referenced, when available.

As specific requirements of each application can determine a different WLS as the optimal choice, there is no single universal WLS ideal for all applications. The purpose of this review is to provide essential data to allow to make an informed WLS selection for a specific application, or to guide defining further R\&D goals needed to make such a choice.
\section{Fundamentals}
\unskip
\subsection{Basic requirements}
A WLS suitable for experimental applications in liquid argon should have the following properties:
\begin{itemize}[leftmargin=*,labelsep=5.8mm]
    \item high absorption cross-section for the short wavelength light (e.g. 128~nm in LAr),
    \item high photoluminescence quantum yield (PLQY), i.e. the ratio between the number of re-emitted photons to the number of absorbed photons,
    \item high Stokes shift, i.e. low overlap between absorption and emission spectra resulting in the WLS being transparent to its own emission wavelength.
\end{itemize}

From a practical point of view, 128~nm light is efficiently absorbed with an absorption length comparable to its wavelength in a vast majority of materials, which makes the first requirement trivially met (notable exceptions include noble gasses and liquids and specialized fluoropolymers).

The PLQY and the Stokes shift requirements are therefore the most critical. Specifically, the WLS performance at the LAr 128~nm excitation wavelength and at 87~K temperature is essential for use in LAr detectors. This is non-trivial, as the PLQY and Stokes shift can depend on both of the wavelength and temperature. The emission spectrum should ideally have a peak in blue or green in order to match the most efficient photosensors available on the market.

Important for dark matter and 0$\nu\beta\beta$-decay detectors, and less so for other applications, is the degree of radiopurity of WLS materials and technical feasibility of implementing them without introducing radioactive isotopes into the detector, which could potentially lead to dangerous backgrounds.

Last but not least, long term stability in cryogenic conditions is a necessary requirement for all LAr applications.

\subsection{Conversion mechanisms}
\begin{figure}[ht]
\centering\includegraphics[width=\linewidth]{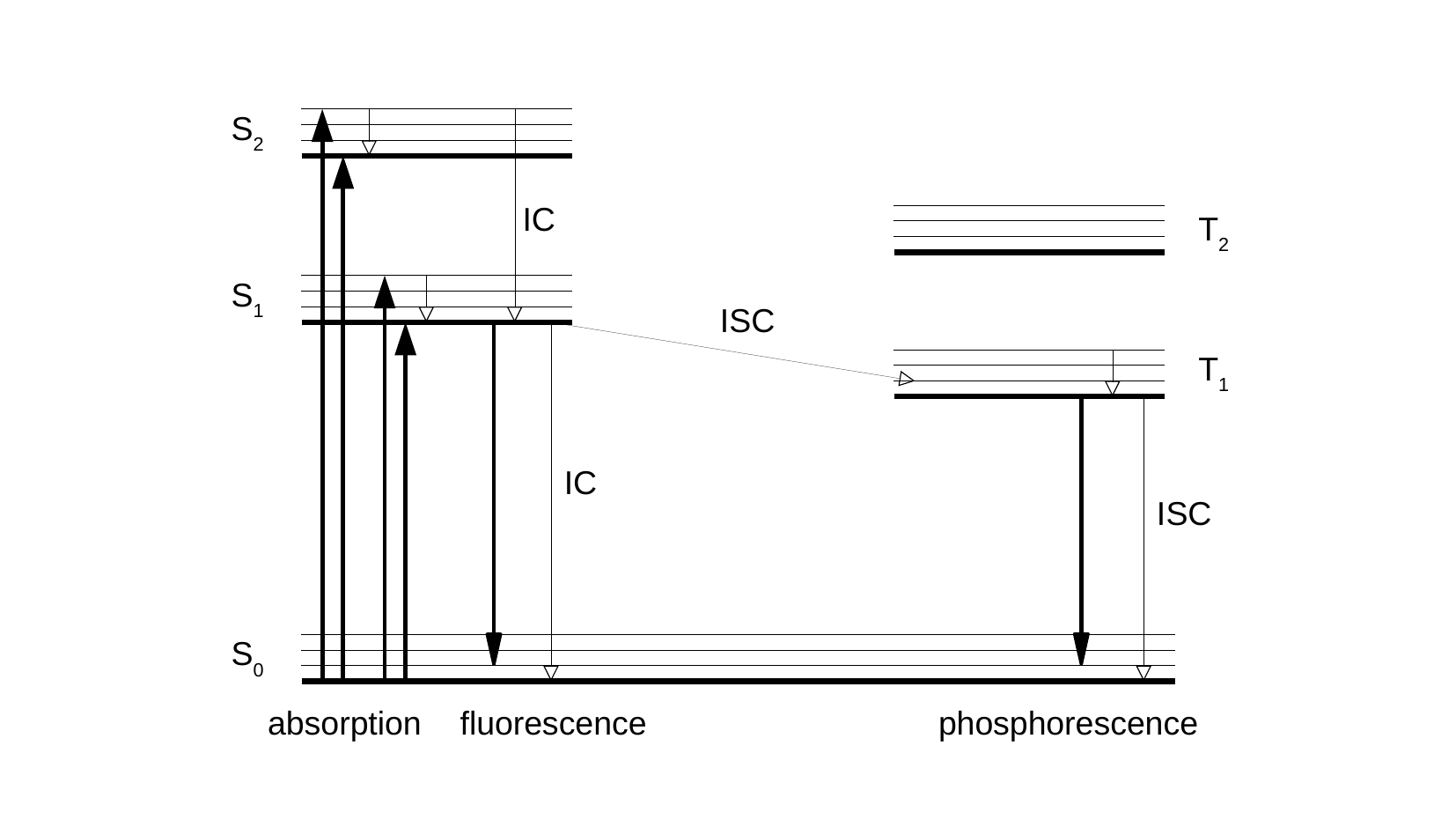}
\caption{Jablonski diagram, explaining photophysical processes in molecular systems. Empty arrows denote non-radiative transitions: vibrational relaxation, internal conversion (IC) and intersystem crossing (ISC), see~\cite{Valeur.ch3}.}
\label{fig:jablonsky}
\end{figure}

Figure~\ref{fig:jablonsky} depicts the relation between main processes which play the role in wavelength shifting: absorption, internal conversion, fluorescence, phosphorescence, intersystem crossing. 
Absorption of a photon brings a molecule from the fundamental electronic state, $S_0$, to one of the vibrational levels of the first singlet electronic state, $S_1$, (timescale: 10$^{-15}$~s), which is then followed by a (non-radiative) vibrational relaxation towards the lowest vibrational level of S1 (10$^{-12}$ - 10$^{-11}$~s). 
Fluorescence is the emission associated with the relaxation $S_1\rightarrow S_0$ (10$^{-10}$ - 10$^{-7}$~s).
Due to the energy loss in the excited state due to vibrational relaxation, the fluorescence spectrum is shifted towards higher wavelengths (lower energy) than the absorption spectrum. The Stokes shift is then the gap between the maximum of the first absorption band and the maximum of fluorescence.

Naively, the fluorescence emission wavelength
should always be longer than that of absorption, however at room temperature some molecules may be in a
vibrational level higher than the lowest level of $S_0$, which could lead to partial overlap of the energy and emission spectra.

The higher energy singlet states such as $S_2$ relax to $S_1$ via non-radiative processes. Radiative transitions predominantly take place from the lowest excited states of $S_1$ to the ground state $S_0$ or one of its excited vibrational sub-levels. In the case of phosphorescence, the process is similar except it orginates from the first electronic triplet state, $T_1$. The internal conversion process typically competes with the radiative transition (fluorescence) but has much lower probability. The transition  $S_1\rightarrow S_0$ is classified as spin-allowed, and is very quick taking on the order of
$\sim$10$^{-9}$~s. On the other hand, the timescale of the $T_1\rightarrow S_0$ transition falls between $\mu$s and ms because the process is spin-forbidden. 
\section{Wavelength shifter types and properties}
There are several WLS used in LAr experiments. In this section we discuss their basic properties and compare them.

Table~\ref{tab:wls} summarizes available experimental data and practical information on the properties of the most representative WLS examples, described in more detail in the rest of this Section. In a number of cases, and particularly for the PLQY parameter for some WLS, significant discrepancies in the published results are present. There are multiple possible reasons for this, which are not always clearly identifiable, including: (1) aging effects or degradation of either the WLS (e.g. TPB) sample or the reference sample, (2) geometrical effects due to inconsistent optical diffusivity of the WLS and reference samples, (3) coating quality (pinholes) and deposition technique details, and (4) uncertainties regarding the intensity of the impinging light.
\begin{table}[H]
\caption{Fundamental properties of common WLS materials used in LAr detectors: peak emission wavelength ($\lambda_{em}$), PLQY, re-emission lifetime ($\tau$),  refractive index ($n$), vapour pressure ($p_{sat}$), and approximate sublimation temperature ($T_m$).}\label{tab:wls}
\centering
\tablesize{\footnotesize}
\begin{tabular}{l|ccccccc}
\toprule
\textbf{Wavelength shifter}	& \textbf{$\lambda_{em}$ [nm]}	& \textbf{PLQY @ 128~nm} & {\bf $\tau$ [ns]} & {\bf n} & {\bf p$_{sat}$ [mbar]} & {\bf $T_m$ [}$^\circ${\bf C]} & {\bf Comment} \\
\midrule
TPB & 430 & 0.6 \cite{Benson}--2 \cite{Graybill:2020xhu} & 2 & 1.7  & 10$^{-11}$ & 204 \\
p-Terphenyl & 350 & 0.82~\cite{Kumar:79} & 1 & 1.65 & & 213 & PLQY @ 254 nm \\
bis-MSB & 440 &	0.75--1~\cite{Gehman_2013, Araujo} & 1.5 & 1.7 &  & 180 & PLQY rel. to TPB \\
pyrene & 470 & 0.64~\cite{Birks} & 155 & 1.8 & 6$\cdot$10$^{-6}$& 150 & PLQY @ 260 nm\\
PEN & 420 &	0.4--0.8~\cite{KuzniakPEN} & 20 & 1.75 & -- & 270 &PLQY rel. to TPB \\
\bottomrule
\end{tabular}
\end{table}

\subsection{Organic wavelength shifters}
The luminescence of organic compounds is essentially based on localized $\pi$-electron systems (i.e. single or conjugated aromatic rings) within individual organic molecules.

Organic molecules with well-developed $\pi$-conjugated systems tend to show intense electronic absorption in the UV to visible wavelength regions because of a transition from the bonding $\pi$* orbital into the anti-bonding $\pi$* orbital (i.e. a transition from $S_0$ to $S_1$ state)~\cite{fundphosphors}.

\begin{figure}[H]
\centering
\includegraphics[width=0.9\linewidth]{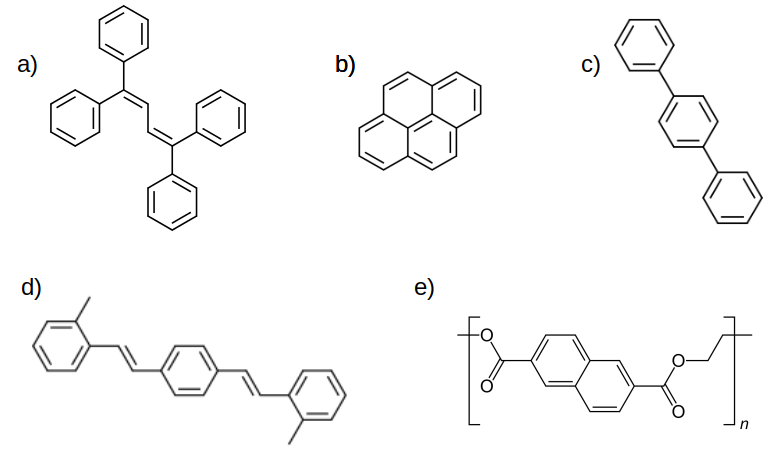}
\caption{Chemical structures of wavelength shifters relevant for liquid argon detectors: (\textbf{a}) 1,1,4,4-tetraphenyl-1,3-butadiene (TPB), (\textbf{b}) pyrene, (\textbf{c}) p-terphenyl, (\textbf{d})  1,4-Bis(2-methylstyryl)benzene (bis-MSB), (\textbf{e}) polyethylene naphthalate (PEN).}\label{fig:organics}.
\end{figure}

\begin{figure}[ht]
\centering\includegraphics[width=\linewidth]{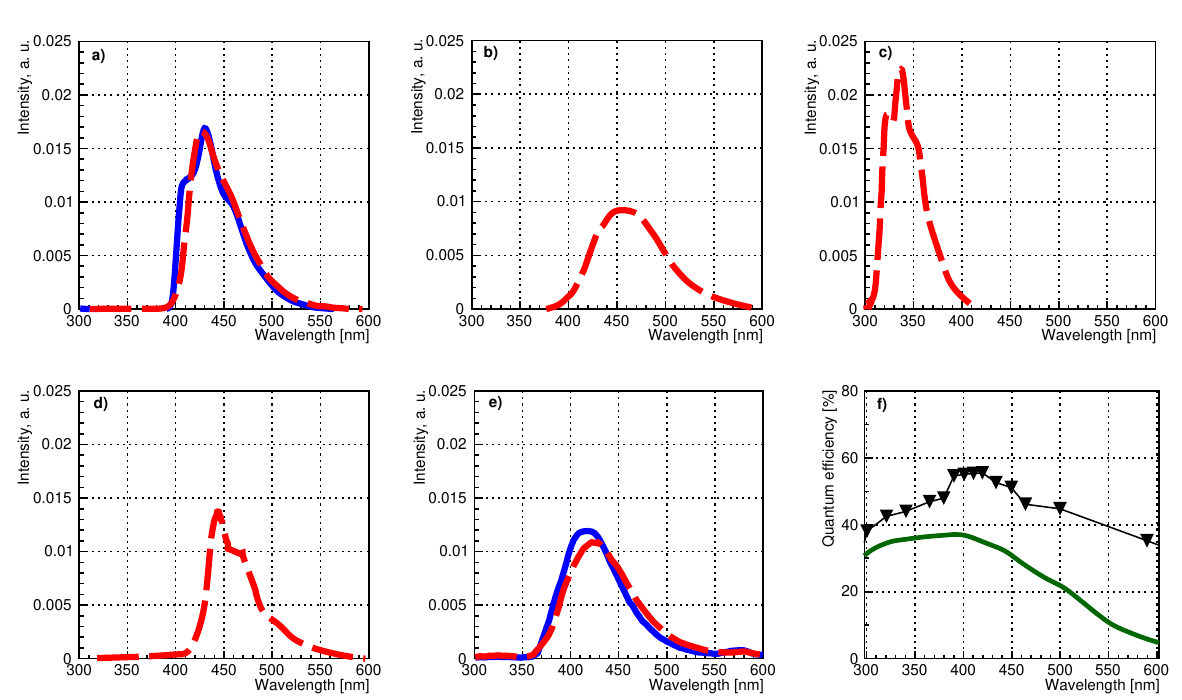}
\caption{Emission spectra of wavelength shifters relevant for liquid argon detectors, area normalized, and quantum efficiency curves of photosensors. Spectra measured at room temperature are shown (red, dashed lines) and, when available, also at 87~K (blue solid lines). (\textbf{a}) 1,1,4,4-tetraphenyl-1,3-butadiene (TPB)~\cite{Francini}, (\textbf{b}) pyrene~\cite{Birks}, (\textbf{c}) p-terphenyl~\cite{ptpspectrum}, (\textbf{d})  1,4-Bis(2-methylstyryl)benzene (bis-MSB)~\cite{Gehman_2013}, (\textbf{e}) polyethylene naphthalate (PEN)~\cite{Mary_1997}. (\textbf{f}) Quantum efficiency curves at room temperature for FBK NUV-HD-SF SiPMs at 10~V overvoltage (black triangles)~\cite{ds20k} and Hamamatsu R11065 PMTs (green solid line)~\cite{Acciarri_2012}.}
\label{fig:spectra}
\end{figure}

The most common class of organic WLS, used in LAr detector applications, are conjugated aromatic hydrocarbons and their derivatives, see chemical structures in Figure~\ref{fig:organics} and the emission spectra in Figure~\ref{fig:spectra}.
Generally, their absorption as well as the emission maxima, shift to longer wavelengths with increasing number of the conjugated aromatic rings.

Although not discussed further in the text, for completeness we highlight other major classes of organic WLS:
\begin{itemize}
    \item Heterocycles ($\pi$-conjugated molecules containing heteroatoms), e.g. coumarine or PPO, are another common class of fluors, however with lower photostability in comparison to aromatic hydrocarbons; and no significant history in the context of LAr detectors.
\item Nanostructured organosilicon luminophores (NOL)~\cite{ineos} are a class of compounds in which aromatic hydrocarbon or heterocycle-based molecules are connected through silicon atoms, which permits excitation energy transfer via the F\"orster mechanism (FRET). NOLs can be designed and optimized for particular properties, including high PLQY at VUV excitation. While currently not used in LAr detectors, their applications in liquid xenon have been explored~\cite{AKIMOV2012403,Akimov_2017}. 
\end{itemize}

In general, the quantum yields of organic WLS fluorescence are strongly dependent on the temperature because of its larger effects on thermal non-radiative decay processes. At LAr temperature quantum yields tend to be high as many rotational and vibrational motions of pendant groups are frozen.

\subsubsection{TPB}\label{sec:tpb}
Because of its robustness and blue emission spectrum conveniently matching the typical PMT photocathode response, TPB is the most commonly used WLS in LAr applications. It was first identified as a WLS with high PLQY (improving by a factor of three with respect to sodium salicylate commonly used at that time) already in 1967~\cite{brunet, mai}, and subsequently proposed as a candidate for commercial and research purposes (UV astronomy)~\cite{Burton:73}. Together with p-terphenyl it was then recognized as a useful WLS for dark matter detectors based on xenon~\cite{tpb2}. McKinsey et al.~\cite{MCKINSEY1997351} reported results for TPB and several other fluors, comparing their PLQY under exposure to gaseous argon scintillation light for coatings deposited using all major techniques used until today (see Section~\ref{sec:coatings}). In particular, for evaporated coatings a broad efficiency maximum was reported for thickness in the 0.2-0.3~mg/cm$^2$ range.

An extensive study of the TPB WLS properties as a function of the excitation wavelength, temperature and coating type has been published by Francini et al.~\cite{Francini}. For coatings evaporated on specular reflector substrates, Vikuiti Enhanced Specular Reflector (ESR) by 3M~\cite{Vikuity,Weber2451}, the vibronic emission spectrum structures of TPB, unresolved at room temperature, emerged at cryogenic temperatures. At least four distinct structures were visible at 87~K, which were attributed to the stretching mode vibration of trans-butadiene. This effect was later independently confirmed~\cite{Corning_2020} and is typical for pure organic films. It is not present in either liquid or solid (e.g. polymeric matrix) solutions, where the randomness of the medium around the TPB molecule leads to broadening of the vibronic structures into a single, unresolved band. At  87~K the 128~nm excitation PLQY increased by about 10\% with respect to the room temperature conditions~\cite{Francini}.

 TPB can form four known polymorphs~\cite{polymorph1,polymorph2} depending on its deposition or growth conditions (temperature, pressure, solvent), as confirmed on vacuum evaporated coatings~\cite{Broerman_2017}. Due to different spatial packing these polymorphs exhibit slightly different solid properties and may also subtly differ in the vibronic structure. Therefore, the exact conditions of the coating production process have impact on the properties of the final WLS film.

Microcrystaline structure of TPB coatings leads to significant scattering of visible light. The mean scattering length for a vacuum evaporated coating was measured to be in the range 2–3~$\mu$m~\cite{Stolp_2016}.

Despite the importance of this parameter and the widespread use of TPB, its absolute PLQY remains controversial. While early TPB measurements used sodium salicylate as a reference, an absolute measurement of PLQY as a function of excitation wavelength down to 120~nm was first attempted in Ref.~\cite{GEHMAN2011116}, which reported a constant shape of the emission spectrum regardless of the wavelength and a PLQY of 1.2 at 128~nm. However, in a subsequent paper~\cite{Benson} released several years later, that result was attributed to miscalibration of the photodiode used. A re-analysis of that original experiment was presented based on (1) the corrected photodiode calibration and (2) a geometric acceptance evaluated with Geant4/RAT Monte Carlo including the full optical model of the coating and of the setup. Both the old re-analyzed data and data from a new series of measurements indicated significantly lower PLQY values of approximately 0.6. As a validation of the optical model used, experimental data on coating response intensity vs. coating thickness were reproduced with the simulation. Finally, Graybill et al.~\cite{Graybill:2020xhu} has recently reported the TPB effective conversion efficiency of approximately 2 at the LAr scintillation wavelength, indicating also significant aging effects that can lead to efficiency decreasing by up to a factor of three.
Clearly, several factors are at play in this puzzle, including: aging effects (discussed in more detail in Section~\ref{sec:aging}), poorly known and controlled absolute PLQY of reference samples used in the historic measurements, and non-trivial light propagation effects caused by scattering in TPB itself and at the optical interfaces with the medium and with the substrate, which are notoriously difficult to model and account for.

Hints of multiphoton emission from TPB are possibly linked with the issue of delayed fluorescence of TPB. 
The dominant fluorescence time constant of TPB for evaporated coatings is short: $\sim$1.7~ns~\cite{FLOURNOY1994349}.
Additional longer time constants, attributed to triplet excited states, have been first reported in alpha induced TPB scintillation~\cite{POLLMANN2011127, Veloce_2016}. These time constants formed a long non-exponential and temperature dependent emission tail, which was approximated and parametrized with a sum of exponentials, with time constants reaching 40~$\mu$s.
Segreto~\cite{SegretoPhysRevC.91.035503, Segreto_2016} reported the presence of similar delayed emission also in the VUV-induced TPB fluorescence, in a measurement using gaseous Ar with quenched triplet emission as the prompt light source, and pointed out that the LAr scintillation photon energy (9.7~eV) likely exceeds the ionization potential of TPB. Electrons released in the process would have enough energy to excite both singlet and triplet states of surrounding molecules. Excited triplet states decay via internal conversion to the lowest lying triplet state $T_1$, which is long-lived. These states interact through the triplet-triplet interaction, $T_1+T_1\rightarrow S_1+S_0$, and at the end the delayed scintillation photon is produced by the de-excitation of the resulting $S_1$ state. Sanford et al.~\cite{PhysRevD.98.062002} followed up with a measurement of TPB response in LAr to alphas, betas and 128~nm photons, using a longer 2~ms acquisition window. The data revealed even longer $\sim$ms scale delayed emission and generally were well described with the Voltz-Laustriat~\cite{VoltzLaustriat} model, developed to describe the triplet-triplet interaction dynamics in scintillators. Such TPB time response model, combined with a LAr scintillation model, which included a LAr physics-driven intermediate component, and the instrumental response, was then used to successfully fit the LAr pulse shapes in the DEAP-3600 detector~\cite{DeapPulse}.

Summarizing the above results, there is convincing evidence for ionization of TPB with 128~nm photons, which leads to the excitation of both singlet and triplet TPB states, with the triplet excitation responsible for the delayed TPB emission. Naively, this could lead to the PLQY of $\sim$2, as observed by Graybill et al.~\cite{Graybill:2020xhu}, however with half of the photons delayed and therefore not useful for modes of detection requiring fast conversion (needed for PSD). At the same time, inhibitors, such as dissolved or trapped oxygen, could affect the triplet, $T_1$, TPB states. e.g. causing quenching~\cite{VoltzLaustriat},\cite[Section A.1.3]{stanford}, further complicating the overall picture. This highlights the need for further tests, in particular for time resolved measurements of various types of coatings.

Recapitalizing, TPB remains the most robust WLS of choice for LAr detectors, and has been characterized in the greatest level of detail, although some of its fundamental properties are still debated in the literature and deserve further research. Looking forward, the main challenge is with the scale-up for applications in the future very large detectors, where both efficient production of many hundreds of square meters of coated surfaces, and long term stability, on the timescale of a decade or more, are crucial.

\subsubsection{Polyethylene naphthalate}\label{sec:pen}
PEN is chemically a cousin of the well known polyethylene terephthalate (PET or Mylar). Because of its low oxygen permeability and convenient mechanical, electrical and optical properties, it is industrially mass produced for applications in e.g. food packaging or electronics. Although it is also one of the components of the Vikuiti ESR, specular reflector commonly used in LAr detectors, and its fluorescence properties have been known for a long time~\cite{Ouchi}, it was only proposed as a scintillator in 2011~\cite{Nakamura_2011}, and then, recently, as a WLS particularly well suited for large scale LAr detectors~\cite{KuzniakPEN}.

The broad deep blue (430~nm) emission band from PEN originates from excimer formation between naphthalene-dicarboxylate units which results in a $^1$($\pi$,$\pi$*) fluorescence transition. Therefore, similar to pyrene and contrary to other discussed organic WLS, it exhibits no vibrational peak structure upon cooldown.

As discussed in~\cite{KuzniakPEN}, early literature suggested a significant increase of PEN PLQY at 128~nm excitation wavelengths and 87~K, compared to room temperature and near UV excitation, leading to extrapolated values consistent with TPB in LAr detector conditions. More recent ex-situ measurements performed at room temperature relative to TPB indicate PLQY in the range of 0.3-0.6 w.r.t. TPB~\cite{KuzniakPEN, Efremenko_2020}. Similar results were obtained in the ProtoDUNE-DP detector where the PEN foils were installed in LAr as WLS in front of a fraction of PMTs~ \cite{SotoOton}. In the context of a significant spread of PLQY results for TPB discussed in Section~\ref{sec:tpb}, it is difficult to interpret these results in absolute terms until measurements against a well calibrated and stable reference will be made. Similarly, R\&D work on PEN scintillation performed in the context of LEGEND, has shown that the scintillation yield of commercially available PEN is lower by a factor of three~\cite{penpos} when compared to the original results by Nakamura et al.~\cite{Nakamura_2011}; a significant spread of WLS yields for various PEN samples has also been recently reported~\cite{ProtoDPspread}. Although the origin of this apparent discrepancy remains unclear, it might be caused by aging effects (similar as in the case of TPB), differences in the production process, or by the different composition of additives between various PEN grades. It is possible that a tailored and optimized PEN production process could lead to an improved yield, comparable to TPB.

The fluorescence lifetime of PEN is approximately 20~ns~\cite{Janecek} at room temperature, based on measurements on Vikuiti ESR. Qualitatively, given that no significant distortion of the LAr scintillation time constants was observed in other time-resolved measurements using either ESR~\cite{Baudis_2015} or PEN~\cite{SotoOton}, the PEN lifetime at 87~K should be of similar value. No  subdominant longer time constants have been reported so far in the photoluminescence spectrum. There is, however, evidence of delayed scintillation emission induced by irradiation with alpha particles~\cite[Section 7.3]{minicleanesr}.

The main advantage of PEN is the ease of use and low cost. It appears that even commercially produced off the shelf grades of PEN perform sufficiently well for use in certain types of LAr detectors. PEN is available in large format rolls or sheets and can be easily used as a liner for large LAr tanks. The maximum PLQY achievable with PEN, assuming optimized production process and appropriate storage and handling conditions, remains an open question.

\subsubsection{p-Terphenyl}
P-terphenyl (pTP) was one of the first WLS studied for applications in noble liquid detectors~\cite{MCKINSEY1997351}. Its PLQY was found to be comparable with TPB, however the broad emission band extending below 350~nm was less convenient for use with common photosensors. Also, lower concentrations of pTP than TPB were achievable in polystyrene (PS) matrix without crystallization effects.

Otherwise, many phenomena known for TPB also apply for pTP, including the evidence for delayed alpha induced scintillation, see~\cite{tpbtpt,tpbtpt0}.

Recently, pTP has found an important application in ARAPUCA modules, which are under development for DUNE, see Section~\ref{sec:arapuca} for details.

pTP is listed as very toxic to aquatic life.

\subsubsection{bis-MSB}
1,4-Bis(2-methylstyryl)benzene (bis-MSB) is a WLS used in large organic liquid scintillator experiments, where it is directly dissolved in the liquid scintillator mixture. Because of its quantum efficiency close to unity~\cite{bismsbchina} at near UV wavelengths, fast re-emission time constant and cost effectiveness, it has also been considered an interesting candidate for the LAr long baseline neutrino program.

References \cite{Gehman_2013, BaptistaMSB_2013} describe relative 128~nm conversion efficiency measurements of polymeric films doped with bis-MSB and TPB, showing comparable yields. A more recent measurement showed consistent PLQY at 128~nm for evaporated coatings of both materials~\cite[Appendix A2]{Araujo}.

One disadvantage of bis-MSB is its toxicity, significantly higher than for other discussed WLS materials, and its environmental impact on aquatic life.
\subsubsection{Pyrene}
Birks et al.~\cite{Birks} described the properties and fluorescence mechanism of pyrene in detail. The fluorescence spectrum has features attributed to excimer (at ~470~nm) and monomer emission (below 400~nm). Although the absorption spectrum of pyrene is characteristic of the monomer, the emission spectrum is consistent with the expectation for excimer emission. The crystal structure of pyrene is dimeric in form -- the elementary unit of the crystal lattice is a pair of parallel pyrene molecules. Monomers initially excited by the incident light can rapidly ($<$1~ns) convert into excimer excitons with $>$100~ns lifetime, where the excitation physically manifests as reduced distance between both molecules. 

The temperature dependence of pyrene fluorescence results in a $\sim~$50\% efficiency and time constant increase at 87 K compared to room temperature~\cite{Birks}.

Due to high vapour pressure, pyrene coatings are unstable in vacuum, and cannot be deposited with vacuum evaporation. Pyrene can be used mixed with polymers, as the movement and diffusion of large fluor molecules in a polymeric matrix is very strongly constrained, as they essentially become trapped between tangled polymeric chains, which leads to some of the lowest diffusion constants ever measured (e.g. D=9.4$\cdot$10$^{-19}$~cm$^2$/s for pyrene in PS~\cite{pyreneinps}). The ratio of excimer to monomer emission increases with pyrene concentration in the polymer matrix.

Because of its long time constant, pyrene can be used for background mitigation purposes, currently under study with potential application in the context of the DEAP-3600 detector, see Section~\ref{sec:bgr}.

Pyrene is listed as very toxic to aquatic life with long lasting effects.

\subsection{Inorganic wavelength shifters}
In solid inorganic fluors luminescence is determined by their lattice structures, as opposed to the molecular structures in organic WLS, which has a few consequences. Firstly, the luminescence of inorganic materials is altered or disappears altogether when the crystals melt or decompose. In particular it is very different for nanoparticles (or quantum dots) and chemically identical macroscopic bulk materials. Secondly, the lattice structure is much more resistant to aging and UV damage effects than organic WLS molecules.

Elementary gases can also be used as WLS; one of the first wavelength shifters used with noble gas scintillators was a gas admixture of nitrogen~\cite{Aprile}, although it also acted as a quencher, which reduced the overall scintillation yield. Currently, vibrant R\&D is underway on employing xenon as a distributed WLS for LAr detectors.

\subsubsection{Nanoparticles}
Application of inorganic nanoparticles in LAr detectors has been proposed recently~\cite{nano1}. Nanoparticle-based WLS coatings deposited directly on photosensors could be used instead of traditional organic WLS. 
A consequence of the small size of particles is the increase of the energy band gap as well as formation of discrete energy levels at the edges of the band gap. The absorption maximum is shifted to higher energies (from visible to UV wavelengths). 

One of the main advantages of inorganic WLS is the durability and resistance to aging effects that, as discussed later, is a common issue for organic WLS materials. In addition, the emission wavelength can be tuned to match the optimal sensitivity of an existing photodetector, since the energy gap is highly dependent on the nanoparticle size.

Ref.~\cite{nano1} lists ZnS:Mn,Eu, ZnS:Mn, Cu-Cy, CdTe and LaF$_3$:Ce nanoparticles as showing the most potential for LAr applications. The most significant photosensor QE enhancement is observed in the 225–375~nm range, with emission maxima ranging between 330 and 600~nm, however also with fairly long re-emission time constants (32~ns for CdTe, significantly more for the rest).

Another study~\cite{perovskite} explored the potential of perovskite (CsPbBr$_3$) quantum dots and showed up to threefold conversion efficiency enhancement when compared to a TPB-coated sensor, with $<$1~ns lifetime emission at $\sim$510~nm, and expected even several times higher PLQY at cryogenic temperatures, which is very promising.

Nanoparticle coatings were deposited from solution through casting, dip or spray-coating, which are techniques scalable to very large surface areas. Adequate radiopurity of such coatings would have to be demonstrated for application in dark matter and 0$\nu\beta\beta$-decay detectors.

\subsubsection{Xenon} \label{sec:xenondope}
R\&D on LAr doped with small admixtures of xenon tested in TPB-coated detectors, hinted that such mixtures could deliver better energy resolution, faster decay time (potentially useful for TOF-PET and medical applications), increased Rayleigh scattering length (helping event position reconstruction) and improved Pulse Shape Discrimination (PSD)~\cite{Peiffer_2008,Lippincott_2014}.

The process of the triplet (long lived) excitation transfer from Ar to Xe occurs through collisions and Xe dimers de-excite through emission at 175~nm, essentially without any efficiency loss. This wavelength is significantly more accessible to conventional photosensors than 128~nm. Akimov et al~\cite{Rudik_2019} showed that LAr doped with even higher concentrations of Xe, and without TPB, has a better PSD efficiency than pure LAr or Xe-doped LAr with TPB. That implied that the singlet (short lived) Ar excitation is transferred to Xe as well, and a mechanism assuming emission of 128~nm photon, subsequently absorbed by Xe, was proposed.

A detailed energy transfer model, together with experimental evidence showing that Xe is a faster WLS than TPB, is available in Ref.~\cite{Galbiati:2020eup}. The WLS character of Xe indicated by the above brings additional benefits for the event position reconstruction: (1) WLS is now distributed in LAr with conversion happening at the event location, more direct information is available through hit patterns, and (2) timing is faster, which enables more precise time-of-flight analysis.

PSD performance on a larger data sample, long term stability of Xe-doped LAr mixtures, including risks and implications of potential aggregation processes, as well as potential Xe concentration gradients over very large detector volumes are subject to further studies.

\section{Coating production techniques}\label{sec:coatings}
\unskip
\subsection{Vacuum evaporation}
In the past vacuum evaporation has been most commonly used for WLS deposition, of TPB in particular. It is based on elevating the temperature of the WLS in powdered or crystalline form up to the sublimation point in 10$^{-5}$~mbar or better vacuum. Resistive or electron gun heating is typically applied, targeting relatively low 150-250$^\circ$C temperatures, which are sufficient for sublimation of organic WLS molecules. In vacuum the mean free path of evaporated molecules is sufficiently long to allow them to reach the substrate and stick to its surface.

A typical process geometry involves a crucible and a flat substrate sheet or plate located 10-50~cm away. Since molecules essentially follow straight trajectories, the distance from the crucible allows to control the large scale uniformity of the coating, see e.g.~\cite{POLLMANN2011127} and references therein. 

Controlled temperature or rotating substrate is sometimes used to improve the small scale uniformity of coatings, however satisfactory results have also been obtained on acrylic and glass substrates without these features.
More complicated substrate geometries are also possible to coat with custom designed crucibles, e.g.:
\begin{itemize}[leftmargin=*,labelsep=5.8mm]
\item on an inner surface of an acrylic tube, with a resistive NiCr mesh wrapped into a long narrow filament, containing pockets of TPB~\cite{deap1,dart}
\item on long optical fibers, with a movable crucible~\cite{Csathy:2016wdy},
\item on long rectangular 120$\times$25~cm$^2$ diffuse reflector sheets, with an array of 13 crucibles~\cite{ArDM_refl_2009},
\item on the inner surface of an 85~cm radius sphere (9~m$^2$ surface area),cryst with a spherical crucible suspended in the center~\cite{Broerman_2017}.
\end{itemize}

Largest campaigns of TPB evaporation on rectangular and planar ESR foils were performed for WArP~\cite{Francini} and SBND~\cite{Szelc_2016} detectors.
The vacuum evaporation technique becomes increasingly challenging for coating large surface areas, mainly due to the high vacuum requirement, and a duty cycle dominated by the pump down periods of the evaporation chamber needed to obtain the desired quality of vacuum.

\subsection{Solvent coating}
Solvent coating is an attempt to avoid scale-up challenges related to vacuum evaporation. The discussed WLS materials (with the exception of PEN) are soluble in common organic solvents, including toluene and dichloromethane, although typically concentrations of only up to several percent are achievable. The solution is then applied to the substrate using one of many available techniques, such as brushing, dipping, spraying/airbrushing (e.g. with ethyl ether as a solvent~\cite{MCKINSEY1997351}) or spin coating. Proportionally large amounts of organic, especially chlorinated, solvents necessary for such processes is a complication on its own due to significant toxicity and environmental impact of such chemicals.

A typical challenge for this class of techniques is achieving uniform and reproducible coatings, which requires substantial effort needed for optimizing and automating the process. Another issue is the crystallization of the WLS once the solvent evaporates, which can result in relatively large crystals, sometimes poorly attached to the substrate, and a non-negligible fraction of the substrate left uncoated. This can directly translate into reduced WLS efficiency and lead to cracking or poor stability of the film when immersed in LAr.

Nevertheless, the process can be optimized for a given substrate and lead to a robust product, as demonstrated in R\&D effort for GERDA~\cite{Baudis_2015}, where a Tetratex substrate coating with a solution of TPB in dichloromethane was developed, with the WLS efficiency about 2 times lower in comparison to evaporated coatings. A porous character of Tetratex was a possible factor allowing the solution to saturate the substrate and, after drying, leave a strongly attached TPB film. Films of 0.9~mg/cm$^2$ thickness were stable for long-term operation in liquid argon and insensitive to exposure to ambient humidity and oxygen.

A demonstration of the spin-coating technique with toluene used as the solvent~\cite{Yang:2019zob}, resulted in approximately 200~nm thick TPB coating (uniform, although with pinholes visible with SEM) showing 0.25 efficiency relative to evaporated TPB coating. 

\subsection{Doped polymeric matrix}
The motivation behind embedding WLS in polymer matrix is twofold: similarly to the solvent method, simplification of the coating production process and stabilizing the WLS against long term aging or degradation effects. 
The choice of the host polymer has consequences; if the polymer matrix, e.g. polymethyl methacrylate (PMMA), has no aromatic fragments, its repeating unit cannot be transferred to an excited state with a lifetime
sufficient for further energy transfer. PS, on the other hand, can take part and mediate in the energy transfer between WLS molecules~\cite[Section 3.3]{Bower:994867}.

For LAr WLS applications the coatings typically made are solid solutions, with WLS molecules not chemically bound to polymeric chains. 
In the vast majority of literature reports polymer based WLS coatings (usually TPB in PS), deposited with a variety of techniques (brushing, spraying, dipping) led to coatings with the conversion efficiency 2-3 times lower than that of evaporated coatings~\cite{Mavrokoridis_2011, baptista2014benchmarking}. The process typically involves dissolving through stirring or shaking finely powdered polymer in the solvent, optionally at elevated temperature, in concentration of the order of 1\% and a similar amount of TPB. Admixture of ethanol reduces the surface tension of the mixture and improves the wettability during e.g. dip-coating~\cite{moss2016factor}.

Excellent results have been achieved with spin-coated TPB/PS films. PS grades with long chains and {\cal O}(100k) molecular weight allowed one to dope larger concentrations of TPB, up to 0.8 TPB/PS mass ratio, without or with only limited aggregation or crystallization effects. Optimum spin-coated TPB/polymer films should use high TPB mass ratio suspensions and 100~nm film thickness~\cite{Graybill:2020xhu}, and can reach up to twice the efficiency of evaporated TPB coatings, however with caveats discussed in Section~\ref{sec:aging}.

As another successful example, nylon foil "minishrouds" coated with 30–40\% TPB in PS were developed by the GERDA collaboration in order to improve light collection and in the vicinity of the Ge detectors, and mitigate Ar-42/K-42 backgrounds~\cite{GerdaEnclosures}. The final concentration was a compromise between the mechanical stability and the intensity of emitted light. 
The coating was applied by brushing both sides of transparent nylon foils with the dichloromethane based solution and, typically, 0.3~mg/cm$^2$ of dry film mass deposited. Various tests were performed in order to understand the usability of such a foil in GERDA. 
No signs of deterioration, flexibility loss or significant difference between spectra collected before and after up to several months exposure to air and to cryogen were observed.

\subsection{Lamination}
PEN, available in large format polymeric foils, is suitable for lamination with reflective or photosensitive substrates~\cite{KuzniakPEN} or other mechanical attachment schemes. Runs have been performed using the CERN FLIC 50 l prototype TPC, with the DUNE-like resistive cathode (1) covered with an ESR foil evaporated with TPB and (2) coated with PEN, with no obvious issues observed~\cite[Section 5.16.2.1]{dune}.

\section{Stability and aging}\label{sec:aging}
\subsection{TPB}
Mechanical stability of WLS films has been a point of concern and a subject of multiple studies. Whether coated in a pure or a doped polymeric form, the WLS layer has to withstand temperature cycling, the transient thermal stresses during the detector cool-down as well as deliver reasonably stable performance over the duration of the experiment, which typically means several years.

Extensive R\&D campaigns were undertaken by several groups to study and find the right match of the deposition method and the substrate providing satisfactory stability and durability, see~\cite{ArDM_refl_2009, Baudis_2015, Broerman_2017}. For vacuum evaporated coatings it has been determined that on the commonly used substrates, such as quartz, ITO-coated quartz, acrylic, or ESR foils, stable and robust coatings are achieved as long as the film growth rate during the deposition process is kept very low, i.e. approximately at $\sim$1~\AA/s. The resulting coatings are possible to scratch, but generally are resistant to gentle wiping, manipulation and thermal cycling. As determined from microscope imaging, rich crystalline structure of the coatings appears to be resistant even to abrupt liquid nitrogen dunk tests, as long as they are performed in dry atmosphere~\cite{Broerman_2017}. Not all substrates behave in the same way, and there are anecdotal reports of reduced resistance to thermal cycling of TPB deposited on ITO-coated plastics. 

It has been found that in liquid xenon after 20~h of immersion the surface density of evaporated TPB films decreased by 1.6~$\mu$g/cm$^2$, although without any morphologic or chemical surface changes detected with SEM micrographs or XPS. This can be interpreted as a slow dissolution process where the TPB molecules remain intact~\cite{xetpb}. Recently, a similar phenomenon, although significantly slower than in xenon, has been observed in LAr~\cite{tpbdissolved}. Evidence has been found for TPB concentration in LAr gradually increasing over a 50~h soak time, up to tens of parts per billion in LAr by mass, for both evaporated TPB and TPB/PS coating samples present in the system. The nature of this effect is a subject of further studies.

Independently from the above effects, aging effects related with environmental and, primarily, UV exposure of TPB are known. Already McKinsey et al.~\cite{MCKINSEY1997351} reported that exposure to the ordinary fluorescent lighting in laboratory air for 24~h resulted in a twofold yield reduction, with similar accounts confirmed in more detail elsewhere~\cite{Acciarri_2013, Chiu_2012}. In particular, the degrading effect of the fluorescence lighting was confirmed, but found to be mitigated through the use of amber filters. No difference in degradation rate between storage in ambient air vs. inert dry gas was observed. This leads to a clear recommendation for storage of WLS films in dark conditions and limiting the exposure to (unfiltered) lighting during handling.

Yahlali et al.~\cite{YAHLALI2017109} demonstrated, however, that the conversion efficiency loss after exposure to VUV light comparable with that integrated over a $\sim$5 year lifetime of a typical experiment does not to exceed 15\%, which is consistent with the results of e.g. GERDA, DEAP-1, DEAP-3600 or DarkSide-50. Also, a thick TPB coating (1.6~$\mu$m) had stronger resistance to VUV radiation damage than a thin coating (130 nm).

The nature of the UV damage to TPB has been studied in detail \cite{Jones_2013}, and was attributed to formation of benzophenone, which has known UV blocking properties, and which accumulates in the surface layer of the coating under exposure to UV light, and over time leads to yellowing of the coatings, caused by its further degradation products. The performance and durability difference between the commercially available standard and the scintillation TPB grades was also attributed to the presence of benzophenone. 

As reported by~\cite{Graybill:2020xhu} and other sources, immersion in the polymeric matrix protects TPB molecules from UV induced degradation effects. However, given that TPB emanation in LAr has been reported also from TPB/PS foils~\cite{tpbdissolved}, long term stability of such very thin  ($\sim$100~nm thick) spin-coated films deserves further studies in-situ.

\subsection{PEN}
Mechanical robustness of PEN, in a form of a polymeric film, in the LAr detector environment is implied by the widespread and successful use of ESR foils for this application, as PEN is one of the two components of ESR~\cite{KuzniakPEN}.

Some commercially available grades of PEN films are subjected to high temperature treatment during production (annealing), which can lead to growth of cyclic oligomers on the surface~\cite{penoligomers}. The cyclic oligomer crystals can be removed by washing the film with methylethylketone (or with sanding). Their presence on significant fraction of the surface could conceivably affect the film WLS efficiency at VUV wavelengths.

There are no known accounts of humidity or storage in ambient air affecting the fluorescence or scintillation properties of PEN; in fact, this material is highly impermeable to both air and water. However, UV-induced degradation effects are known. Up to an order of magnitude deterioration of PLQY for excitation wavelengths in the 260--380~nm range as well as a shift of the peak emission wavelength towards higher wavelengths after 5~h of exposure to light from a 30~W broadband UVA lamp has been reported~\cite{penaging}. Qualitatively, the magnitude of this effect appears comparable with the case of TPB, with measured scintillation and fluorescence yields up to 3 times lower than the best ones reported in the literature. Similar to the case of TPB, there is a significant spread of PLQY values reported in the literature (as discussed in Section~\ref{sec:pen}), which might be caused by inconsistent levels of UV degradation in tested samples. (As an additional complication, in most cases PEN yield characterization has been performed relative to TPB coatings, which introduces additional dependence on the TPB reference history and aging effects).

While the UV degradation effects deserve further studies, there is good motivation for storage of PEN films in dark conditions, similar to TPB.

\subsection{p-Terphenyl} 
Similar to the case of TPB, there is evidence of liquid xenon dissolving pTP coatings, evaporated on PTFE reflectors and the entrance window of a photomultiplier tube~\cite{ptpsolubility}.

\subsection{Pyrene}
Pyrene is particularly susceptible to oxidation, and therefore has to be applied, handled and stored in oxygen free conditions. Oxidation leads to yellowing and reduction of both the PLQY and the fluorescence lifetime. The effect is attributed to the formation of a surface layer of pyrenequinone~\cite{Birks}.

\section{Examples of Application}\label{sec:applications}
\unskip
\subsection{Scintillation light collection}
WLS materials are used in LAr experiments both as a transmissive coating on, or in front of, the photon detectors, and as reflective coating on passive surfaces to maximize light collection.

Different types of experiments use different general arrangements of WLS and photon detectors.
Dark matter detectors generally opt for high photosensor coverage and augment them with WLS reflectors covering the walls as well as the gaps between photosensors. The importance of this reflector coverage grows particularly as collaborations move towards the next generation experiments with surface areas exceeding hundreds of m$^2$. We note that the the overall light yield ($LY$) is strongly dependent on their reflectivity ($R$), because of the large number of wall reflections before encountering a photosensor: $LY \sim PLQY \cdot R^n$, where $n$ is the inverse of the photosensor coverage fraction (e.g., for 20\% photosensor coverage, n=5). This underlines the importance of using reflectors with optimized conversion efficiency and reflectivity. 
In the case of neutrino detectors the photosensors are usually placed only behind the transparent anode planes with sparse photosensor coverage. As this configuration leads to reduced light yield and non-uniformities in neutrino detector response~\cite{Garcia-Gamez:2020xrv}, 
WLS reflectors partly covering the walls or cathode have been used by SBND~\cite{Szelc_2016} and LArIAT~\cite{lariat} experiments. 
Another possible configuration is to distribute the WLS throughout the liquid argon volume. This solution requires finding a WLS that uniformly mixes with the liquid argon, e.g. by being soluble. An example is xenon doping of LAr, described in Sec.~\ref{sec:xenondope}, which results in a large fraction of the scintillation light being emitted at the 178~nm characteristic for xenon.


Large-scale neutrino detectors use the liquid argon scintillation light as a trigger (MicroBooNE, SBND) to determine the interaction time, $t_0$ needed to reconstruct the absolute position in the drift coordinate, to remove out-of-beam cosmic events, tag decays as well as improve timing and energy reconstruction. For these applications, low detection efficiency on the order of 10$^{-3}$ is sufficient~\cite{Szelc_2016,Sorel_2014}.

The enhanced collection efficiency obtained in dark matter experiment geometries leads to an increased light yield and, in consequence, a lower energy threshold.
The energy threshold is a parameter particularly important for dark matter detectors, as the expected spectrum of nuclear recoils induced by elastic scattering of weakly interacting massive particles (WIMPs) increases steeply towards low energies. Therefore the sensitivity of WIMP detectors strongly and non-linearly depends on the energy threshold and  maximizing the PLQY and minimizing optical losses is of crucial importance.
In the DEAP-3600 detector~\cite{deap} as much as 16\% of the original number of scintillation photons are registered as photoelectrons; increasing this number even further is mandatory for the next generation detectors which will not be attainable without efficient use of WLS.

There is a broad range of WLS applications in detectors where the LY requirements fall somewhere in-between the accelerator neutrino and dark matter search experiments. These less stringent light yield requirements of, e.g., 0$\nu\beta\beta$-decay detectors combined with the technical challenges of the vacuum evaporation technique motivated R\&D on alternative coating methods. These passive elements are often combined with sparse arrays of photosensors (as in the LAr veto for the DarkSide-20k detector~\cite{argo}, for which PEN is considered a candidate WLS) and/or optimized light collection schemes allowing to at least partially recover the response uniformity, e.g. by collecting light from a large area in the detector and guiding it to a limited number of photosensors (as in GERDA or DUNE). 

As collaborations approach the construction of future generation large-scale detectors, such as Argo~\cite{argo}, developing a scalable WLS solution allowing to maintain high LY and response uniformity is crucially important and is currently a very active R\&D topic, which has  strong synergies with the large-scale neutrino experiments, particularly, DUNE~\cite{dune}.

The WLS applications in LAr detectors can be split into passive elements of the photon detection systems, such as WLS coated reflective elements, simple coated photon-sensors, photon-sensors with two-stage WLS and photon traps. The WLS coatings used can be produced in a variety of ways, as discussed in Section~\ref{sec:coatings}. Their thickness is usually optimized for the specific geometry or place of application, i.e. typically thinner coatings are used in front on optical windows or photosensors and thicker ones on reflectors. For example for the most commonly used evaporated TPB coatings the thickness recommended for the transmission geometry was 0.1-0.3~mg/cm$^2$~\cite{Benson}, while thicker coatings yield better results on diffuse or specular reflectors~\cite{ardm}.

\subsubsection{WLS-coated passive elements}

Vacuum evaporated TPB coatings on passive detector elements have been used for the applications where light collection has been most critical, namely for dark matter detectors such as WArP~\cite{warp}, MiniCLEAN~\cite{miniclean}, ArDM~\cite{ArDM_refl_2009}, DEAP-1~\cite{deap1} and DarkSide-50~\cite{ds-50}. The coatings are typically evaporated ex-situ directly on the reflective material, and then used to line the walls of the LAr chamber. Following successful tests in the LArIAT test beam experiment~\cite{lariat}, where the walls and subsequently the cathode were lined with TPB coated ESR foils, the SBND neutrino detector~\cite{Machado:2019oxb} will install TPB-coated ESR foils on the detector cathode which required the largest production of such foils to date: roughly 40~m$^2$ of TPB coated foils were produced. Because of only partial coverage, the resulting light yields are relatively modest compared to dark matter detectors. A production of a similar scale process is planned for the DarkSide-20k TPC. PEN or evaporatively TPB-coated ESR foils are also one of the alternative options considered for the future DUNE modules, see \cite[Section 5.16.2.1]{dune} and~\cite{wlsLOI}.

In the DEAP-3600 detector, the evaporation on the 9~m$^2$ inner surface of the spherical acrylic vessel was performed in-situ, after deploying the evaporation crucible in the center, pumping down the detector chamber, and effectively turning the inner detector volume into a large spherical evaporation chamber~\cite{Broerman_2017}.

In the LAr veto tank of the GERDA detector ESR reflector foils, dip-coated in a toluene based solution of TPB and PS, were used as a liner. After significant R\&D effort these were replaced with a film of pure TPB ($\sim$0.9~mg/cm$^2$ thick) deposited on Tetratex diffuse reflector sheets (PTFE based) via dip coating in a solution of TPB in dichloromethane~\cite{Baudis_2015}. The TPB/Tetratex reflector gives approximately two times more light than the TPB-doped PS on ESR. Additionally, TPB/PS-coated \cite{GerdaEnclosures} nylon foil ''minishrouds'' improving light collection in the vicinity of the Ge detectors were developed.

\subsubsection{WLS-coated photon detectors}

Most liquid argon experiments will opt to coat their photon detectors in WLS, either via direct coating with TPB using evaporative coatings (ArDM,  SBND and ICARUS~\cite{icarustpb}), dip-coatings e.g. TPB combined with polysterene (WArP, LArIAT), by coating a window before the actual PMT (DEAP-1, MiniCLEAN, DarkSide-50) or by using a WLS front plate (MicroBooNE~\cite{ microboone}, ProtodDUNE-DP~\cite{ProtoDPspread}).
In the last case, because of rather low resulting light yield, the scintillation light is primarily used for triggering and identifying high-energy cosmic-ray tracks. In DEAP-3600 the PMTs were not coated with WLS instead relying on the surface of the inner chamber to convert the light which was then transported to the PMTs via light guides.

In large scale detectors the photosensor coverage can be increased by using light collectors, e.g. based on deploying light guides coated with TPB-doped PS~\cite{Bugel_2011}. The configuration which was developed and deployed in ProtoDUNE-SP \cite{ProtoDUNErun}, uses 207.4~cm long PMMA light guides, with  attenuation length on the order of 2~m~\cite{moss2016factor}. Other solutions combine different WLS compounds to enhance the collection efficiency and will be discussed next.

\subsubsection{Two-stage WLS}

A more complex approach involves two stages of wavelength shifting. Firstly, VUV photons are converted to blue wavelengths using e.g. a TPB-based coating. These blue photons are then wavelength shifted again in a polymeric substrate doped with a commercial grade green fluor (e.g. in the form of fibers, plates or light guides), with the absorption spectrum well matched with the emission spectrum of TPB. Due to a more favorable difference of refractive indices between the light guide material and LAr, green light at $\sim$490~nm is more efficiently trapped in the light guide through total internal reflection and benefits from a higher attenuation length in the base material.

An example of such an implementation is the so-called double-shift lightguide bars designed for DUNE and ProtoDUNE-SP. They consist of an outer acrylic plate spray-coated with TPB dissolved in dichloromethane. The blue photons are then absorbed and wavelength-shifted by commercial Eljen EJ-280 plates which are coupled with SiPM detectors~\cite{HOWARD20189, Howard_2018}, yielding a photon detection efficiency of between 0.2\% and 0.48\%. This design was one of the technologies deployed and tested during the ProtoDUNE-SP run~\cite{ProtoDUNErun}.

A concept using a ``curtain'' (50\% coverage) of evaporatively TPB-coated green wavelength shifting fibers (Saint-Gobain BCF-91A), additionally improving the yield and response uniformity of TPB-coated reflectors covering the outer walls of the cryostat, has been developed for GERDA Phase~II~\cite{Csathy:2016wdy, gerdaupgrade} and is planned for LEGEND~\cite{legend}.

A bundle of Kuraray fibers, was installed around the acrylic neck of the DEAP-3600 detector (i.e. outside the LAr volume, contrary to the cases discussed above) to improve light collection of TPB converted light from the region of the detector where light collection was reduced due to a gap in the PMT coverage~\cite{deap}.

\subsubsection{Photon traps}\label{sec:arapuca}
Another concept is to use dichroic filters combined with WLS to trap light and effectively enlarge the effective surface area instrumented with SiPM sensors. This concept is used in the recently developed ARAPUCA~\cite{arapuca},  X-ARAPUCA~\cite{xarapuca}  and ArCLight~\cite{arclight} photon detectors.

The ARAPUCA concept, originally developed for DUNE and installed and tested during the ProtoDUNE-SP run~\cite{ProtoDUNErun} uses a box with one or two faces containing a dichroic filter. The filter is coated with pTP on the outside which converts the light to a wavelength at which the filter-window is transparent. Once the photons cross the filter they encounter a second layer of WLS, in this case TPB, which shifts them to blue, making them reflect off of the filter and effectively trapping them inside of the reflective box until they can find the SiPM detector coupled to it. During the ProtoDUNE-SP run the ARAPUCA were found to achieve a detection efficiency of close to 2\%. The X-ARAPUCA concept improves on the ARAPUCA design by adding a WLS light-guide inside the box which couples to the SiPMs. 
Simulations and preliminary measurements lead to expect a detection efficiency of about 3.5\% for this design. The X-ARAPUCA is the baseline DUNE photon detector as described in the Technical Design Report~\cite[Section 5]{dune}. They will also be installed in SBND, with half of the detectors coated with pTP and half left un-coated to be sensitive only to the light wavelength-shifted and reflected off the cathode.

The ArCLight detectors use a similar concept to the X-ARAPUCA detectors and have been developed for the ArgonCube detector~\cite{auger2019new}, where they are set to be installed in the fieldcage walls. The ArCLight detectors consist of an Eljen  EJ-280 WLS plate covered on the back and sides with highly reflective dielectric foils. The front face is a dichroic filter coated with TPB on the outside. Similarly to the X-ARAPUCA design the photons are first shifted to blue, where they pass the filter and are subsequently shifted to green by the WLS plate which then performs as a lightguide transporting the photons to the coupled SiPMs. The ArCLight detectors have been found to have a detection efficiency of up to 2.2\%~\cite{arclight}.  

\subsection{Impact of WLS on dark matter background mitigation}\label{sec:bgr}
\subsubsection{WLS effects in Pulse Shape Discrimination}

LAr-based dark matter experiments use pulse shape discrimination (PSD)~\cite{Boulay:2006mb} to select nuclear recoils associated with potential WIMP signals and to reject electron recoils typically associated with backgrounds. The difference in ionization density of the two classes of events results in the scintillation signals from the former being dominated by the scintillation from the LAr singlet with a lifetime of~$\sim$7~ns~\cite{hitachi} and those resulting from electron-recoils to be dominated by the slow, LAr triplet, light with a lifetime of $\sim$1.5~$\mu s$. This method has been shown to reject electron-recoil backgrounds at a level of 1:10$^{10}$~\cite{AMAUDRUZ20161}.

The efficiency of the PSD methods depends on the light yield of the detector. The higher the light yield, the more efficient the background discrimination can be at lower recoil energies. Therefore all avenues using WLS to improve scintillation light collection described in the previous section, trivially result in improved background mitigation prospects. 

One should note that the WLS time constants can affect the observed distribution of the detected light. In order for the PSD methods to be effective it is therefore important for the WLS dominant time constant to preserve the original timing structure of the LAr scintillation to a sufficient degree, i.e. it should not be much higher than the LAr singlet lifetime ($\sim$7~ns). This condition is fulfilled by all WLS materials discussed in the review except for pyrene, which will be discussed later in this section. 

We note however, that the TPB emission time~\cite{SegretoPhysRevC.91.035503} combined with LAr scintillation microphysics~\cite{DeapPulse} have been suggested as a culprit for the so-called intermediate time component with a lifetime of $\sim$40~ns observed in a few measurements~\cite{Acciarri_2010}.

\subsubsection{Alpha Background Removal Using WLS}

Apart from gamma/beta backgrounds mitigated with LAr PSD, alpha activity is another potential source of WIMP-like events in, primarily, single-phase dark matter detectors, such as DEAP-3600. 
Radioactive daughters of Rn-222 are particularly problematic, as gaseous radon present in the ambient air can diffuse into the surface layer of materials, leaving behind its radioactive decay daughters, including long lived Pb-210 and alpha decaying Po-210. Topologies of such Po-210 alpha decays in the vicinity of the detector surface, in which most of the alpha energy is deposited in the inactive substrate, or in which only the recoiling nucleus deposits energy in LAr, can lead to apparent-low energy events contaminating the region of interest for dark matter searches~\cite{deap1}. In dual-phase detectors, such as DarkSide-50, surface events are easier to identify, but can still lead to background events through pile-up with other classes of events~\cite{PhysRevD.98.102006}.

Delayed emission from alpha scintillation in TPB and pTP, which is more abundant than for VUV photon conversion, has been researched as a potential handle to reject such events, in addition to e.g position reconstruction and fiducialization. First evidence for such emission, and its low-temperature parameters, together with a simple PSD parameter designed to reject such events has been published~\cite{POLLMANN2011127, Veloce_2016}. 
Further studies in LAr conditions led to an the estimate  that the  surface  background  could  be suppressed by a factor of $\sim$10 (100) for TPB (pTP) at 50\% (90\%) nuclear recoil acceptance~\cite{tpbtpt0} and that for the case of TPB the projected PSD factor could be as high as 10$^3$, when using milliseconds-long integration windows~\cite{PhysRevD.98.062002, tpbtpt}.
In practice, the above mechanism has found so far only limited use in dark matter search analyses~\cite{boqianPhD}. Typically, applying cuts based on position reconstruction is sufficient to reject these surface events. Also, instrumental effects, especially PMT afterpulsing, reduce signal acceptance of cuts based on alpha PSD parameters, unless the acquisition window significantly exceeds the $\sim$10~$\mu$s, typically used in single-phase LAr experiments, which leads to a significant increase in dead time.

\begin{figure}[H]
\centering
\includegraphics[width=0.95\linewidth]{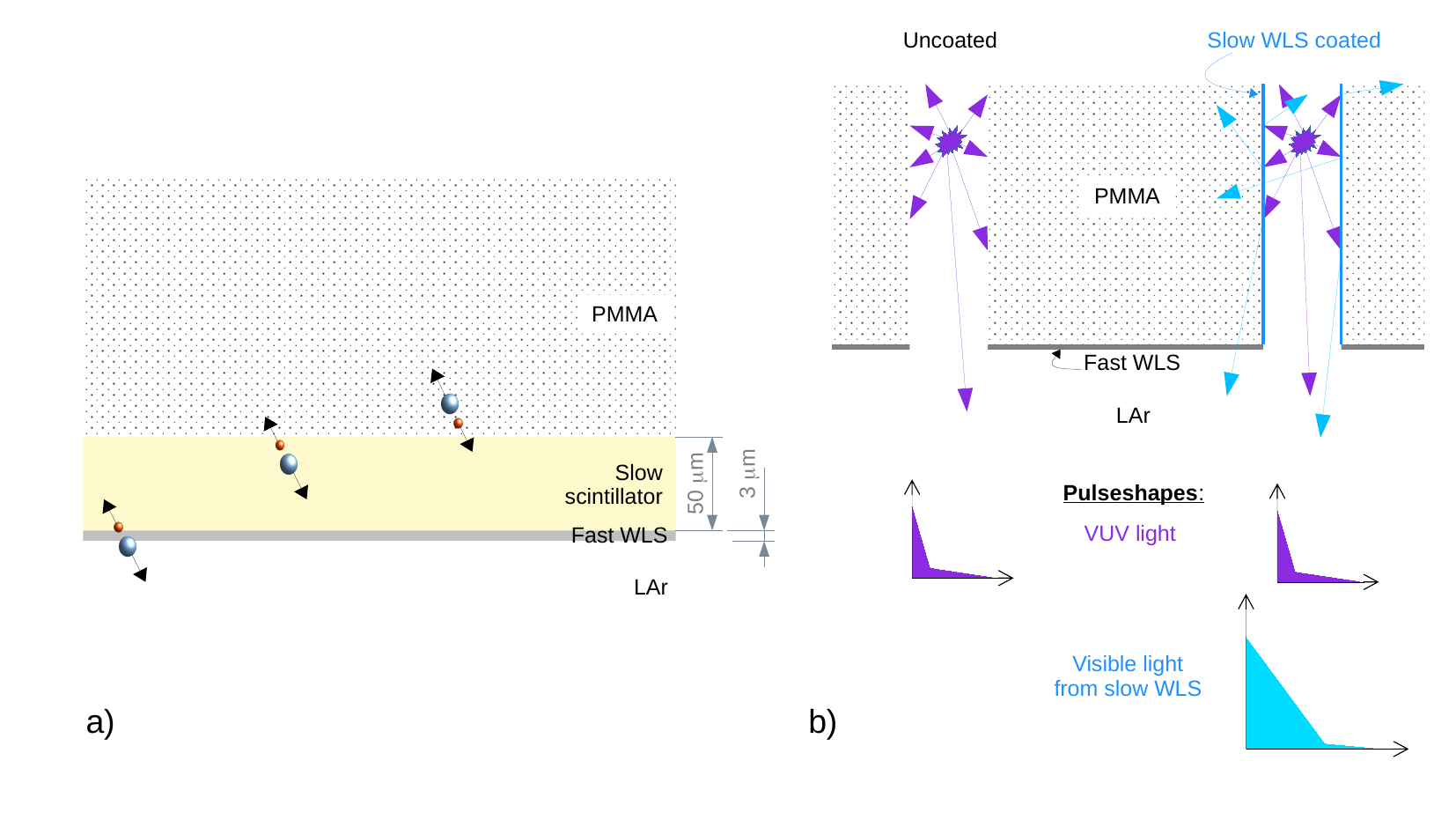}
\caption{WLS-based background mitigation schemes: (\textbf{a}) WLS-coated slow scintillator on a passive PMMA detector wall. Alpha decays will either deposit all of their energy in an active region (LAr, WLS or the scintillator) and be discriminated based on high deposited energy, or will deposit only a fraction of their energy in the slow scintillator, in which case they can be discriminated against based on its long decay time~\cite{surfacepsd}. (\textbf{b}) Slow WLS coating. Alpha decays occurring in a slit in the detector wall can induce a fake low-energy background event, as most of the VUV scintillation is absorbed before entering the main volume (left). In presence of the slow WLS in the slit (right), the event will be dominated by visible photons converted in the slow WLS and could be removed with PSD.}\label{fig:bgr}.
\end{figure}

To maximize the efficiency of surface event PSD rejection in a 10~$\mu$s time window, employing a layer of scintillator with a $>\sim$300~ns emission time constant between the conventional WLS coating, e.g. TPB based and a passive transparent detector wall material (e.g. acrylic~\cite{surfacepsd, surfacepsd2}) has been proposed. This configuration would allow to achieve up to 10$^8$ suppression factor using PSD. The scintillator layer should be sufficiently thick to stop alpha decays originating in the substrate (>$\sim$50~$\mu$m), see Figure~\ref{fig:bgr}(a). This concept could be implemented using e.g. a commercially available slow scintillator material, e.g. Bicron BC-444.

In addition to background events induced by degraded alphas on the detector surfaces, shadowed alpha decays are another class of related backgrounds, as shown in Figure~\ref{fig:bgr}(b). VUV LAr scintillation light from alpha decays occurring in peripheral and shadowed regions of detectors (slits, gaps, inlets/outlets) is mostly absorbed by peripheral surfaces without the WLS coating. In consequence, out of 2$\cdot$10$^5$ initial VUV photons generated by a Po-210 alpha, depending on the solid angle, only small fraction will be detected, leading to a continuum in the apparent energy spectrum extending all the way down to the low energy dark matter search window.

Coating such problematic blind-spots with a slow time constant WLS, substantially alters the timing of such pathological events enabling the use of PSD to solve this problem. Care must be taken to only use such material in the shadowed regions of the detector, so that it would not significantly interfere with timing of events from the main volume of the detector, spoiling the PSD against beta/gamma events. 

Because of its long $\sim$115~ns singlet fluorescence lifetime, pyrene has been identified as a promising candidate for such application. It is primarily coated as a dopant in a PMMA or other polymeric matrix~\cite{captalk}, given its incompatibility with vacuum depositions. At 87~K the time constant of the pyrene excimer fluorescence further increases to approximately 155~ns, which provides a strong PSD handle against events dominantly wavelength shifted in such coating. A polymer based pyrene coating is under investigation for robustness in assembly and operation in the neck of the DEAP-3600 detector, in order to mitigate the alpha backgrounds, which currently limit the sensitivity of the experiment~\cite{deap231}.

In summary, fluorescence timing properties of WLS materials play a very important role in mitigating alpha backgrounds in dark matter searches, and can be employed in a variety of scenarios, useful also in the next generation experiments.


\section{Summary and future directions}

Liquid argon scintillation light is a crucial component of a wide range of detectors - from those searching for low energy nuclear recoils indicating potential dark matter interactions to those detecting GeV-scale accelerator neutrinos. 
This scintillation light can only be feasibly detected with the help of WLS compounds, which are therefore a crucial element of these detectors. The last decade has brought huge progress in understanding the behaviour, performance and stability of the most commonly used WLS compounds, as well as developing new methods of applying them to detect scintillation light.

In parallel there has been exciting development both in identifying new WLS compounds and solutions, as well as innovative applications of known WLS, that provide better or new functionality. However, significant room for further research and development still remains in the field. Future large-scale detectors, such as Argo and DUNE, would require WLS coating of extremely large surfaces. This would be difficult to achieve on a reasonable timescale with the current evaporation technology used for TPB coatings. However, the spin-coating technique could provide a solution, should the high PLQY be confirmed.

PEN, the main alternative to large-scale TPB coatings, which should be relatively easy to produce for large areas, is currently found to be less performant than TPB in terms of PLQY, and would need R\&D to improve that to be used in future dark matter detectors. In addition, all WLS solutions installed in next generation LAr detectors will need to have their long-term stability established at levels much higher than was needed until now. New directions include ideas such as doping the large LAr detectors with xenon, which shifts the light to wavelengths where it is significantly simpler to detect, e.g. not requiring an external pTP coating on the X-ARAPUCA detectors as well as completely new ideas, such as quantum dots.

The importance of R\&D on WLS compounds in noble liquid detectors has been recognized by the recently published report on "Basic Research Needs for High Energy Physics Detector Research \& Development" which lists increased light collection as one of the key elements of Priority Research Direction (PRD) nr 5. The report recognizes the need to collect light "over many tens to hundreds of square meters"~\cite{doe}. It also mentions developing "high-efficiency wavelength-shifting techniques" as an important area of research in PRD 5 and PRD 24, listing several different avenues of proceeding towards achieving this goal.

All of the above-mentioned R\&D areas would benefit greatly from a multidisciplinary approach providing results that would be applicable to very different research areas such as searches for dark matter and 0$\nu\beta\beta$-decay, as well as accelerator neutrino measurements, simultaneously. The LIDINE (LIght Detection In Noble Elements) conference series provides an excellent venue to exchange these kind of ideas and cross-pollinate between the different research fields. Indeed many of the solutions listed in this work were first proposed at one of the conferences from this series. One key need that would greatly help multi-disciplinarity is for facilities that could provide a 'complete' characterization of various WLS solutions. One of the first tasks of such a facility would likely be to resolve the outstanding TPB PLQY puzzle which would require a cryogenic setup, capable of time resolved measurements and an integrating sphere providing absolute calibration.

In this decade, as several extremely large LAr experiments will be constructed and start taking data, the importance of WLS and their applications in LAr-based detectors will only increase. 

\vspace{6pt} 

\authorcontributions{writing--original draft preparation and editing, M.K; writing--sections on examples of application and future directions, review and editing, A.M.S and M.K.}
\funding{This work was supported by the International Research Agenda Programme AstroCeNT (MAB/2018/7) funded by the Foundation for Polish Science (FNP) from the European Regional Development Fund. This project has received funding from the European Union’s Horizon 2020 research and innovation programme under grant agreement No 952480. It was also partially funded by the Science and Technology Facilities Council (STFC), part of the United Kingdom Research and Innovation; and by the Royal Society UK awards: RGF\textbackslash EA\textbackslash 180209 and UF140089.}


\conflictsofinterest{The authors declare no conflict of interest.} 

\abbreviations{The following abbreviations are used in this manuscript:\\

\noindent 
\begin{tabular}{@{}ll}
0$\nu\beta\beta$-decay & Neutrinoless double beta decay \\
bis-MSB & 1,4-Bis(2-methylstyryl)benzene \\
ESR & Enhanced specular reflector \\
FRET & F\"orster resonance energy transfer \\
LAr & Liquid argon \\
NOL & Nanostructured organosilicone luminophore \\
PEN & Polyethylene naphthalate \\
PLQY & Photoluminescence quantum yield \\
PMMA & Poly(methyl methacrylate) \\
PMT & Photomultiplier tube \\
pTP & p-Terphenyl \\
PS & Polystyrene \\
PSD & Pulse shape discrimination \\
SiPM & Silicon photomultiplier \\
TPB & 1,1,4,4-Tetraphenyl-1,3-butadiene \\
WLS & Wavelength shifter \\
\end{tabular}}

\appendixtitles{no} 



\reftitle{References}

\externalbibliography{yes}
\bibliography{larwlsreview}

\begin{thebibliography}{-------}
\providecommand{\natexlab}[1]{#1}

\bibitem[Acciarri and et~al(2010)]{warp}
Acciarri, R.; et~al.
\newblock The {WArP} experiment.
\newblock {\em J. Phys. Conf. Ser.} {\bf 2010}, {\em 203},~012006.
\newblock
  doi:{\changeurlcolor{black}\href{https://doi.org/10.1088/1742-6596/203/1/012006}{\detokenize{10.1088/1742-6596/203/1/012006}}}.

\bibitem[{Amaudruz, P.-A} and et~al(2019)]{deap}
{Amaudruz, P.-A}.; et~al.
\newblock Design and construction of the DEAP-3600 dark matter detector.
\newblock {\em Astropart. Phys.} {\bf 2019}, {\em 108},~1 -- 23.

\bibitem[Rielage and et~al(2015)]{miniclean}
Rielage, K.; et~al.
\newblock Update on the MiniCLEAN Dark Matter Experiment.
\newblock {\em Phys. Procedia} {\bf 2015}, {\em 61},~144 -- 152.
\newblock 13th International Conference on Topics in Astroparticle and
  Underground Physics, TAUP 2013,
  doi:{\changeurlcolor{black}\href{https://doi.org/10.1016/j.phpro.2014.12.024}{\detokenize{10.1016/j.phpro.2014.12.024}}}.

\bibitem[Boccone and et~al(2009)]{ardm}
Boccone, V.; et~al.
\newblock Development of wavelength shifter coated reflectors for the {ArDM}
  argon dark matter detector.
\newblock {\em J. Instrum.} {\bf 2009}, {\em 4},~P06001--P06001.
\newblock
  doi:{\changeurlcolor{black}\href{https://doi.org/10.1088/1748-0221/4/06/p06001}{\detokenize{10.1088/1748-0221/4/06/p06001}}}.

\bibitem[Aalseth and et~al(2018)]{ds20k}
Aalseth, C.E.; et~al.
\newblock DarkSide-20k: A 20 tonne two-phase LAr TPC for direct dark matter
  detection at LNGS.
\newblock {\em Eur. Phys. J. Plus} {\bf 2018}, {\em 133}.
\newblock
  doi:{\changeurlcolor{black}\href{https://doi.org/10.1140/epjp/i2018-11973-4}{\detokenize{10.1140/epjp/i2018-11973-4}}}.

\bibitem[Arneodo and et~al(1997)]{icarus}
Arneodo, F.; et~al.
\newblock The liquid argon TPC for the ICARUS experiment.
\newblock {\em Nucl. Phys. B Proc. Suppl.} {\bf 1997}, {\em 54},~95 -- 104.
\newblock
  doi:{\changeurlcolor{black}\href{https://doi.org/10.1016/S0920-5632(97)00098-4}{\detokenize{10.1016/S0920-5632(97)00098-4}}}.

\bibitem[Acciarri and et~al(2017)]{microboone}
Acciarri, R.; et~al.
\newblock {Design and Construction of the MicroBooNE Detector}.
\newblock {\em J. Instrum.} {\bf 2017}, {\em 12},~P02017,
  \href{http://arxiv.org/abs/1612.05824}{{\normalfont
  [arXiv:physics.ins-det/1612.05824]}}.
\newblock
  doi:{\changeurlcolor{black}\href{https://doi.org/10.1088/1748-0221/12/02/P02017}{\detokenize{10.1088/1748-0221/12/02/P02017}}}.

\bibitem[Szelc(2013)]{lariat}
Szelc, A.M.
\newblock The {LArIAT} light readout system.
\newblock {\em J. Instrum.} {\bf 2013}, {\em 8},~C09011--C09011.
\newblock
  doi:{\changeurlcolor{black}\href{https://doi.org/10.1088/1748-0221/8/09/c09011}{\detokenize{10.1088/1748-0221/8/09/c09011}}}.

\bibitem[Szelc(2016)]{Szelc_2016}
Szelc, A.
\newblock Developing {LAr} scintillation light collection ideas in the Short
  Baseline Neutrino Detector.
\newblock {\em J. Instrum.} {\bf 2016}, {\em 11},~C02018--C02018.
\newblock
  doi:{\changeurlcolor{black}\href{https://doi.org/10.1088/1748-0221/11/02/c02018}{\detokenize{10.1088/1748-0221/11/02/c02018}}}.

\bibitem[Pietropaolo(2017)]{protodune}
Pietropaolo, F.
\newblock Review of Liquid-Argon Detectors Development at the {CERN} Neutrino
  Platform.
\newblock {\em J. Phys. Conf. Ser.} {\bf 2017}, {\em 888},~012038.
\newblock
  doi:{\changeurlcolor{black}\href{https://doi.org/10.1088/1742-6596/888/1/012038}{\detokenize{10.1088/1742-6596/888/1/012038}}}.

\bibitem[Abi and et~al(2020)]{dune}
Abi, B.; et~al.
\newblock Volume {IV}. The {DUNE} far detector single-phase technology.
\newblock {\em J. Instrum.} {\bf 2020}, {\em 15},~T08010--T08010.
\newblock
  doi:{\changeurlcolor{black}\href{https://doi.org/10.1088/1748-0221/15/08/t08010}{\detokenize{10.1088/1748-0221/15/08/t08010}}}.

\bibitem[Agostini \em{et~al.}(2017)Agostini, Bakalyarov, Balata, Barabanov,
  Baudis, Bauer, Bellotti, Belogurov, Belyaev, Benato, Bettini, Bezrukov, Bode,
  Borowicz, Brudanin, Brugnera, Caldwell, Cattadori, Chernogorov, and
  Zuzel]{gerdaupgrade}
Agostini, M.; Bakalyarov, A.; Balata, M.; Barabanov, I.; Baudis, L.; Bauer, C.;
  Bellotti, E.; Belogurov, S.; Belyaev, S.; Benato, G.; Bettini, A.; Bezrukov,
  L.; Bode, T.; Borowicz, D.; Brudanin, V.; Brugnera, R.; Caldwell, A.;
  Cattadori, C.M.; Chernogorov, A.; Zuzel, G.
\newblock Upgrade for Phase II of the Gerda experiment.
\newblock {\em EPJ C} {\bf 2017}, {\em 78},~388.
\newblock
  doi:{\changeurlcolor{black}\href{https://doi.org/10.1140/epjc/s10052-018-5812-2}{\detokenize{10.1140/epjc/s10052-018-5812-2}}}.

\bibitem[Zsigmond(2020)]{legend}
Zsigmond, A.J.
\newblock {LEGEND: The future of neutrinoless double-beta decay search with
  germanium detectors}.
\newblock {\em J. Phys. Conf. Ser.} {\bf 2020}, {\em 1468},~012111.
\newblock
  doi:{\changeurlcolor{black}\href{https://doi.org/10.1088/1742-6596/1468/1/012111}{\detokenize{10.1088/1742-6596/1468/1/012111}}}.

\bibitem[Tayloe(2018)]{Tayloe_2018}
Tayloe, R.
\newblock The {CENNS}-10 liquid argon detector to measure {CEvNS} at the
  Spallation Neutron Source.
\newblock {\em J. Instrum.} {\bf 2018}, {\em 13},~C04005--C04005.
\newblock
  doi:{\changeurlcolor{black}\href{https://doi.org/10.1088/1748-0221/13/04/c04005}{\detokenize{10.1088/1748-0221/13/04/c04005}}}.

\bibitem[3dp(2018)]{3dpi}
3D$\pi$ Meeting, GSSI,  2018.
\newblock \url{https://indico.gssi.it/event/7}.

\bibitem[Nikkel \em{et~al.}(2012)Nikkel, Gozani, Brown, Kwong, McKinsey, Shin,
  Kane, Gary, and Firestone]{Nikkel_2012}
Nikkel, J.A.; Gozani, T.; Brown, C.; Kwong, J.; McKinsey, D.N.; Shin, Y.; Kane,
  S.; Gary, C.; Firestone, M.
\newblock Liquefied Noble Gas ({LNG}) detectors for detection of nuclear
  materials.
\newblock {\em J. Instrum.} {\bf 2012}, {\em 7},~C03007--C03007.
\newblock
  doi:{\changeurlcolor{black}\href{https://doi.org/10.1088/1748-0221/7/03/c03007}{\detokenize{10.1088/1748-0221/7/03/c03007}}}.

\bibitem[Erlandson and et~al(2019)]{alarm}
Erlandson, A.; et~al.
\newblock {ALARM: A mini Liquid Argon Radiation Monitor },  2019.
\newblock Presented at LIDINE 2019,
  \url{https://indico.hep.manchester.ac.uk/contributionDisplay.py?contribId=14\&sessionId=7\&confId=5456}.

\bibitem[Aprile \em{et~al.}(2006)Aprile, Bolotnikov, Bolozdynya, and
  Doke]{Aprile}
Aprile, E.; Bolotnikov, A.E.; Bolozdynya, A.I.; Doke, T., Technology of Noble
  Gas Detectors.
\newblock In {\em Noble Gas Detectors}; John Wiley \& Sons, Ltd,  2006;
  chapter~8, pp. 239--276.
\newblock
  doi:{\changeurlcolor{black}\href{https://doi.org/10.1002/9783527610020.ch8}{\detokenize{10.1002/9783527610020.ch8}}}.

\bibitem[Chepel and Ara{\'{u}}jo(2013)]{Chepel_2013}
Chepel, V.; Ara{\'{u}}jo, H.
\newblock Liquid noble gas detectors for low energy particle physics.
\newblock {\em J. Instrum.} {\bf 2013}, {\em 8},~R04001--R04001.
\newblock
  doi:{\changeurlcolor{black}\href{https://doi.org/10.1088/1748-0221/8/04/r04001}{\detokenize{10.1088/1748-0221/8/04/r04001}}}.

\bibitem[Marchionni(2013)]{marchionni}
Marchionni, A.
\newblock Status and New Ideas Regarding Liquid Argon Detectors.
\newblock {\em Annu. Rev. Nucl. Part. Sci.} {\bf 2013}, {\em 63},~269--290.
\newblock
  doi:{\changeurlcolor{black}\href{https://doi.org/10.1146/annurev.nucl.012809.104445}{\detokenize{10.1146/annurev.nucl.012809.104445}}}.

\bibitem[Gallina and et~al(2019)]{vuv4}
Gallina, G.; et~al.
\newblock Characterization of the Hamamatsu VUV4 MPPCs for nEXO.
\newblock {\em Nucl. Instrum. Methods Phys. Res. A} {\bf 2019}, {\em 940},~371
  -- 379.
\newblock
  doi:{\changeurlcolor{black}\href{https://doi.org/10.1016/j.nima.2019.05.096}{\detokenize{10.1016/j.nima.2019.05.096}}}.

\bibitem[Zheng \em{et~al.}(2020)Zheng, Jia, and Huang]{vuvpd}
Zheng, W.; Jia, L.; Huang, F.
\newblock Vacuum-Ultraviolet Photon Detections.
\newblock {\em iScience} {\bf 2020}, {\em 23},~101145.
\newblock
  doi:{\changeurlcolor{black}\href{https://doi.org/10.1016/j.isci.2020.101145}{\detokenize{10.1016/j.isci.2020.101145}}}.

\bibitem[{The Global Argon Dark Matter Collaboration}(2018)]{argo}
{The Global Argon Dark Matter Collaboration}.
\newblock Future Dark Matter Searches with Low-Radioactivity Argon,  2018.
\newblock {Input to the European Particle Physics Strategy Update 2018-2020},
  \url{https://indico.cern.ch/event/765096/contributions/3295671}.

\bibitem[Valeur and Berberan-Santos(2012)]{Valeur.ch3}
Valeur, B.; Berberan-Santos, M.N., Characteristics of Fluorescence Emission.
\newblock In {\em Molecular Fluorescence}; John Wiley \& Sons, Ltd,  2012;
  chapter~3, pp. 53--74.
\newblock
  doi:{\changeurlcolor{black}\href{https://doi.org/10.1002/9783527650002.ch3}{\detokenize{10.1002/9783527650002.ch3}}}.

\bibitem[Benson \em{et~al.}(2018)Benson, Orebi~Gann, and Gehman]{Benson}
Benson, C.; Orebi~Gann, G.; Gehman, V.
\newblock {Measurements of the intrinsic quantum efficiency and absorption
  length of tetraphenyl butadiene thin films in the vacuum ultraviolet regime}.
\newblock {\em EPJ C} {\bf 2018}, {\em 78},~329.
\newblock
  doi:{\changeurlcolor{black}\href{https://doi.org/10.1140/s10052-018-5807-z}{\detokenize{10.1140/s10052-018-5807-z}}}.

\bibitem[Graybill \em{et~al.}(2020)Graybill, Shahi, Coplan, Thompson, Vest, and
  Clark]{Graybill:2020xhu}
Graybill, J.R.; Shahi, C.B.; Coplan, M.A.; Thompson, A.K.; Vest, R.E.; Clark,
  C.W.
\newblock {Extreme ultraviolet photon conversion efficiency of tetraphenyl
  butadiene}.
\newblock {\em Appl. Opt.} {\bf 2020}, {\em 59},~1217--1224.
\newblock
  doi:{\changeurlcolor{black}\href{https://doi.org/10.1364/AO.380185}{\detokenize{10.1364/AO.380185}}}.

\bibitem[Kumar and Datta(1979)]{Kumar:79}
Kumar, V.; Datta, A.K.
\newblock Vacuum ultraviolet scintillators: sodium salicylate and p-terphenyl.
\newblock {\em Appl. Opt.} {\bf 1979}, {\em 18},~1414--1417.
\newblock
  doi:{\changeurlcolor{black}\href{https://doi.org/10.1364/AO.18.001414}{\detokenize{10.1364/AO.18.001414}}}.

\bibitem[Gehman(2013)]{Gehman_2013}
Gehman, V.M.
\newblock {WLS} R{\&}D for the detection of noble gas scintillation at {LBL}:
  seeing the light from neutrinos, to dark matter, to double beta decay.
\newblock {\em J. Instrum.} {\bf 2013}, {\em 8},~C09007--C09007.
\newblock
  doi:{\changeurlcolor{black}\href{https://doi.org/10.1088/1748-0221/8/09/c09007}{\detokenize{10.1088/1748-0221/8/09/c09007}}}.

\bibitem[Araujo(2019)]{Araujo}
Araujo, G.R.
\newblock Wavelength Shifting and Photon Detection of Scintillation Light from
  Liquid Argon.
\newblock Master's thesis, Technische Universitit\"at M\"unchen,  2019.
\newblock
  \url{http://deap3600.ca/wp-content/uploads/2019/07/MastersThesis\_GRAraujo\_27\_03.pdf}.

\bibitem[Birks \em{et~al.}(1966)Birks, Kazzaz, King, and Flowers]{Birks}
Birks, J.B.; Kazzaz, A.A.; King, T.A.; Flowers, B.H.
\newblock 'Excimer' fluorescence - IX. Lifetime studies of pyrene crystals.
\newblock {\em Proc. R. Soc. Lond. A} {\bf 1966}, {\em 291},~556--569.
\newblock
  doi:{\changeurlcolor{black}\href{https://doi.org/10.1098/rspa.1966.0114}{\detokenize{10.1098/rspa.1966.0114}}}.

\bibitem[{Ku\'zniak, M} \em{et~al.}(2019){Ku\'zniak, M}, {Broerman, B},
  {Pollmann, T}, and {Araujo, G. R}]{KuzniakPEN}
{Ku\'zniak, M}.; {Broerman, B}.; {Pollmann, T}.; {Araujo, G. R}.
\newblock Polyethylene naphthalate film as a wavelength shifter in liquid argon
  detectors.
\newblock {\em Eur. Phys. J. C} {\bf 2019}, {\em 79},~291.
\newblock
  doi:{\changeurlcolor{black}\href{https://doi.org/10.1140/epjc/s10052-019-6810-8}{\detokenize{10.1140/epjc/s10052-019-6810-8}}}.

\bibitem[Yen \em{et~al.}(2006)Yen, Shionoya, and Yamamoto]{fundphosphors}
Yen, W.; Shionoya, S.; Yamamoto, H.
\newblock {\em Fundamentals of phosphors}; Taylor \& Francis Group,  2006.

\bibitem[Borshchev and et~al.(2019)]{ineos}
Borshchev, O.V.; et~al..
\newblock Highly Efficient wavelength shifters: design, properties, and
  applications.
\newblock {\em INEOS OPEN} {\bf 2019}, {\em 2},~112--123.
\newblock
  doi:{\changeurlcolor{black}\href{https://doi.org/10.32931/io1916r}{\detokenize{10.32931/io1916r}}}.

\bibitem[Akimov \em{et~al.}(2012)Akimov, Akindinov, Alexandrov, Belov,
  Borshchev, Burenkov, Danilov, Kovalenko, Luponosov, Ponomarenko, Stekhanov,
  Surin, Zav'yalov, and Yablokov]{AKIMOV2012403}
Akimov, D.; Akindinov, A.; Alexandrov, I.; Belov, V.; Borshchev, O.; Burenkov,
  A.; Danilov, M.; Kovalenko, A.; Luponosov, Y.; Ponomarenko, S.; Stekhanov,
  V.; Surin, N.; Zav'yalov, S.; Yablokov, M.
\newblock Development of VUV wavelength shifter for the use with a visible
  light photodetector in noble gas filled detectors.
\newblock {\em Nucl. Instrum. Methods Phys. Res. A} {\bf 2012}, {\em 695},~403
  -- 406.
\newblock New Developments in Photodetection NDIP11,
  doi:{\changeurlcolor{black}\href{https://doi.org/10.1016/j.nima.2011.12.036}{\detokenize{10.1016/j.nima.2011.12.036}}}.

\bibitem[Akimov \em{et~al.}(2017)Akimov, Belov, Borshchev, Burenkov, Grishkin,
  Karelin, Kuchenkov, Martemiyanov, Ponomarenko, Simakov, Stekhanov, Surin,
  Timoshin, and Zeldovich]{Akimov_2017}
Akimov, D.; Belov, V.; Borshchev, O.; Burenkov, A.; Grishkin, Y.; Karelin, A.;
  Kuchenkov, A.; Martemiyanov, A.; Ponomarenko, S.; Simakov, G.; Stekhanov, V.;
  Surin, N.; Timoshin, V.; Zeldovich, O.
\newblock Test of {SensL} {SiPM} coated with {NOL}-1 wavelength shifter in
  liquid xenon.
\newblock {\em J. Instrum.} {\bf 2017}, {\em 12},~P05014--P05014.
\newblock
  doi:{\changeurlcolor{black}\href{https://doi.org/10.1088/1748-0221/12/05/p05014}{\detokenize{10.1088/1748-0221/12/05/p05014}}}.

\bibitem[Francini \em{et~al.}(2013)Francini, Montereali, Nichelatti, Vincenti,
  Canci, Segreto, Cavanna, Pompeo, Carbonara, Fiorillo, and Perfetto]{Francini}
Francini, R.; Montereali, R.M.; Nichelatti, E.; Vincenti, M.A.; Canci, N.;
  Segreto, E.; Cavanna, F.; Pompeo, F.D.; Carbonara, F.; Fiorillo, G.;
  Perfetto, F.
\newblock {VUV}-Vis optical characterization of Tetraphenyl-butadiene films on
  glass and specular reflector substrates from room to liquid Argon
  temperature.
\newblock {\em J. Instrum.} {\bf 2013}, {\em 8},~P09006--P09006.
\newblock
  doi:{\changeurlcolor{black}\href{https://doi.org/10.1088/1748-0221/8/09/p09006}{\detokenize{10.1088/1748-0221/8/09/p09006}}}.

\bibitem[{Akgun} \em{et~al.}(2008){Akgun}, {Bilki}, {Albayrak}, {Bruecken},
  {Cankocak}, {Clarida}, {Cremaldi}, {Duru}, {Moeller}, {Mestvirishvili},
  {Onel}, {Ozok}, {Sanders}, {Sonmez}, {Wetzel}, {Winn}, and
  {Yetkin}]{ptpspectrum}
{Akgun}, U.; {Bilki}, B.; {Albayrak}, E.A.; {Bruecken}, P.; {Cankocak}, K.;
  {Clarida}, W.; {Cremaldi}, L.; {Duru}, F.; {Moeller}, A.; {Mestvirishvili},
  A.; {Onel}, Y.; {Ozok}, F.; {Sanders}, D.; {Sonmez}, N.; {Wetzel}, J.;
  {Winn}, D.; {Yetkin}, T.
\newblock P-Terphenyl deposited quartz plate calorimeter prototype.
\newblock  2008 IEEE Nuclear Science Symposium Conference Record,  2008, pp.
  2228--2233.
\newblock
  doi:{\changeurlcolor{black}\href{https://doi.org/10.1109/NSSMIC.2008.4774796}{\detokenize{10.1109/NSSMIC.2008.4774796}}}.

\bibitem[Mary \em{et~al.}(1997)Mary, Albertini, and Laurent]{Mary_1997}
Mary, D.; Albertini, M.; Laurent, C.
\newblock Understanding optical emissions from electrically stressed insulating
  polymers: electroluminescence in poly(ethylene terephthalate) and
  poly(ethylene 2,6-naphthalate) films.
\newblock {\em J. Phys. D Appl. Phys.} {\bf 1997}, {\em 30},~171--184.
\newblock
  doi:{\changeurlcolor{black}\href{https://doi.org/10.1088/0022-3727/30/2/004}{\detokenize{10.1088/0022-3727/30/2/004}}}.

\bibitem[Acciarri and et~al(2012)]{Acciarri_2012}
Acciarri, R.; et~al.
\newblock Demonstration and comparison of photomultiplier tubes at liquid Argon
  temperature.
\newblock {\em J. Instrum.} {\bf 2012}, {\em 7},~P01016--P01016.
\newblock
  doi:{\changeurlcolor{black}\href{https://doi.org/10.1088/1748-0221/7/01/p01016}{\detokenize{10.1088/1748-0221/7/01/p01016}}}.

\bibitem[Brunet \em{et~al.}(1963)Brunet, Cantin, Julliot, and Vasseur]{brunet}
Brunet, M.; Cantin, M.; Julliot, C.; Vasseur, J.
\newblock Propri\'et\'es de photocathodes m\'etalliques et de couches
  fluorescentes dans l'ultraviolet lointain.
\newblock {\em J. Phys. Phys. Appl.} {\bf 1963}, {\em 24},~53--59.
\newblock
  doi:{\changeurlcolor{black}\href{https://doi.org/10.1051/jphysap:0196300240305300}{\detokenize{10.1051/jphysap:0196300240305300}}}.

\bibitem[Mai and Drouin(1971)]{mai}
Mai, T.T.H.; Drouin, R.
\newblock Relative Quantum Efficiencies of Some Ultraviolet Scintillators.
\newblock {\em Appl. Opt.} {\bf 1971}, {\em 10},~207--208.
\newblock
  doi:{\changeurlcolor{black}\href{https://doi.org/10.1364/AO.10.0207_1}{\detokenize{10.1364/AO.10.0207_1}}}.

\bibitem[Burton and Powell(1973)]{Burton:73}
Burton, W.M.; Powell, B.A.
\newblock Fluorescence of Tetraphenyl-Butadiene in the Vacuum Ultraviolet.
\newblock {\em Appl. Opt.} {\bf 1973}, {\em 12},~87--89.
\newblock
  doi:{\changeurlcolor{black}\href{https://doi.org/10.1364/AO.12.000087}{\detokenize{10.1364/AO.12.000087}}}.

\bibitem[Lally \em{et~al.}(1996)Lally, Davies, Jones, and Smith]{tpb2}
Lally, C.H.; Davies, G.J.; Jones, W.G.; Smith, N.J.T.
\newblock UV quantum efficiencies of organic fluors.
\newblock {\em Nucl. Instrum. Methods Phys. Res. B} {\bf 1996}, {\em 117},~421
  -- 427.
\newblock
  doi:{\changeurlcolor{black}\href{https://doi.org/10.1016/0168-583X(96)00318-7}{\detokenize{10.1016/0168-583X(96)00318-7}}}.

\bibitem[McKinsey \em{et~al.}(1997)McKinsey, Brome, Butterworth, Golub,
  Habicht, Huffman, Lamoreaux, Mattoni, and Doyle]{MCKINSEY1997351}
McKinsey, D.; Brome, C.; Butterworth, J.; Golub, R.; Habicht, K.; Huffman, P.;
  Lamoreaux, S.; Mattoni, C.; Doyle, J.
\newblock Fluorescence efficiencies of thin scintillating films in the extreme
  ultraviolet spectral region.
\newblock {\em Nucl. Instrum. Methods Phys. Res. B} {\bf 1997}, {\em 132},~351
  -- 358.
\newblock
  doi:{\changeurlcolor{black}\href{https://doi.org/10.1016/S0168-583X(97)00409-6}{\detokenize{10.1016/S0168-583X(97)00409-6}}}.

\bibitem[Vik(2018)]{Vikuity}
Application Guidelines ESR Family,  2018.
\newblock
  \url{https://multimedia.3m.com/mws/media/1389248O/application-guide-for-esr.pdf}.

\bibitem[Weber \em{et~al.}(2000)Weber, Stover, Gilbert, Nevitt, and
  Ouderkirk]{Weber2451}
Weber, M.F.; Stover, C.A.; Gilbert, L.R.; Nevitt, T.J.; Ouderkirk, A.J.
\newblock Giant Birefringent Optics in Multilayer Polymer Mirrors.
\newblock {\em Science} {\bf 2000}, {\em 287},~2451--2456.
\newblock
  doi:{\changeurlcolor{black}\href{https://doi.org/10.1126/science.287.5462.2451}{\detokenize{10.1126/science.287.5462.2451}}}.

\bibitem[Corning \em{et~al.}(2020)Corning, Araujo, Stefano, Pereymak, Pollmann,
  and Skensved]{Corning_2020}
Corning, J.; Araujo, G.; Stefano, P.D.; Pereymak, V.; Pollmann, T.; Skensved,
  P.
\newblock Temperature-dependent fluorescence emission spectra of acrylic
  ({PMMA}) and tetraphenyl butadiene ({TPB}) excited with {UV} light.
\newblock {\em J. Instrum.} {\bf 2020}, {\em 15},~C03046--C03046.
\newblock
  doi:{\changeurlcolor{black}\href{https://doi.org/10.1088/1748-0221/15/03/c03046}{\detokenize{10.1088/1748-0221/15/03/c03046}}}.

\bibitem[Girlando \em{et~al.}(2010)Girlando, Ianelli, Bilotti, Brillante,
  Valle, Venuti, Campione, Mora, Silvestri, Spearman, and Tavazzi]{polymorph1}
Girlando, A.; Ianelli, S.; Bilotti, I.; Brillante, A.; Valle, R.D.; Venuti, E.;
  Campione, M.; Mora, S.; Silvestri, L.; Spearman, P.; Tavazzi, S.
\newblock Spectroscopic and Structural Characterization of Two Polymorphs of
  1,1,4,4-Tetraphenyl-1,3-butadiene.
\newblock {\em Cryst. Growth Des.} {\bf 2010}, {\em 10},~2752--2758.
\newblock
  doi:{\changeurlcolor{black}\href{https://doi.org/10.1021/cg100253k}{\detokenize{10.1021/cg100253k}}}.

\bibitem[Bacchi \em{et~al.}(2014)Bacchi, Brillante, Crocco, Chierotti,
  Della~Valle, Girlando, Masino, Pelagatti, and Venuti]{polymorph2}
Bacchi, A.; Brillante, A.; Crocco, D.; Chierotti, M.R.; Della~Valle, R.G.;
  Girlando, A.; Masino, M.; Pelagatti, P.; Venuti, E.
\newblock Exploration of the polymorph landscape for
  1{,}1{,}4{,}4-tetraphenyl-1{,}3-butadiene.
\newblock {\em CrystEngComm} {\bf 2014}, {\em 16},~8205--8213.
\newblock
  doi:{\changeurlcolor{black}\href{https://doi.org/10.1039/C4CE01046A}{\detokenize{10.1039/C4CE01046A}}}.

\bibitem[Broerman \em{et~al.}(2017)Broerman, Boulay, Cai, Cranshaw, Dering,
  Florian, Gagnon, Giampa, Gilmour, Hearns, Kezwer, Ku{\'{z}}niak, Pollmann,
  and Ward]{Broerman_2017}
Broerman, B.; Boulay, M.; Cai, B.; Cranshaw, D.; Dering, K.; Florian, S.;
  Gagnon, R.; Giampa, P.; Gilmour, C.; Hearns, C.; Kezwer, J.; Ku{\'{z}}niak,
  M.; Pollmann, T.; Ward, M.
\newblock Application of the {TPB} Wavelength Shifter to the {DEAP}-3600
  Spherical Acrylic Vessel Inner Surface.
\newblock {\em J. Instrum.} {\bf 2017}, {\em 12},~P04017--P04017.
\newblock
  doi:{\changeurlcolor{black}\href{https://doi.org/10.1088/1748-0221/12/04/p04017}{\detokenize{10.1088/1748-0221/12/04/p04017}}}.

\bibitem[Stolp \em{et~al.}(2016)Stolp, Dalager, Dhaliwal, Godfrey, Irving,
  Kazkaz, Manalaysay, Neher, Stephenson, and Tripathi]{Stolp_2016}
Stolp, D.; Dalager, O.; Dhaliwal, N.; Godfrey, B.; Irving, M.; Kazkaz, K.;
  Manalaysay, A.; Neher, C.; Stephenson, S.; Tripathi, M.
\newblock An estimation of photon scattering length in tetraphenyl-butadiene.
\newblock {\em J. Instrum.} {\bf 2016}, {\em 11},~C03025--C03025.
\newblock
  doi:{\changeurlcolor{black}\href{https://doi.org/10.1088/1748-0221/11/03/c03025}{\detokenize{10.1088/1748-0221/11/03/c03025}}}.

\bibitem[Gehman \em{et~al.}(2011)Gehman, Seibert, Rielage, Hime, Sun, Mei,
  Maassen, and Moore]{GEHMAN2011116}
Gehman, V.; Seibert, S.; Rielage, K.; Hime, A.; Sun, Y.; Mei, D.M.; Maassen,
  J.; Moore, D.
\newblock Fluorescence efficiency and visible re-emission spectrum of
  tetraphenyl butadiene films at extreme ultraviolet wavelengths.
\newblock {\em Nucl. Instrum. Methods Phys. Res. A} {\bf 2011}, {\em 654},~116
  -- 121.
\newblock
  doi:{\changeurlcolor{black}\href{https://doi.org/10.1016/j.nima.2011.06.088}{\detokenize{10.1016/j.nima.2011.06.088}}}.

\bibitem[Flournoy \em{et~al.}(1994)Flournoy, Berlman, Rickborn, and
  Harrison]{FLOURNOY1994349}
Flournoy, J.; Berlman, I.; Rickborn, B.; Harrison, R.
\newblock Substituted tetraphenylbutadienes as fast scintillator solutes.
\newblock {\em Nucl. Instrum. Methods Phys. Res. A} {\bf 1994}, {\em 351},~349
  -- 358.
\newblock
  doi:{\changeurlcolor{black}\href{https://doi.org/10.1016/0168-9002(94)91363-3}{\detokenize{10.1016/0168-9002(94)91363-3}}}.

\bibitem[Pollmann \em{et~al.}(2011)Pollmann, Boulay, and
  Kuźniak]{POLLMANN2011127}
Pollmann, T.; Boulay, M.; Kuźniak, M.
\newblock Scintillation of thin tetraphenyl butadiene films under alpha
  particle excitation.
\newblock {\em Nucl. Instrum. Methods Phys. Res. A} {\bf 2011}, {\em 635},~127
  -- 130.
\newblock
  doi:{\changeurlcolor{black}\href{https://doi.org/10.1016/j.nima.2011.01.045}{\detokenize{10.1016/j.nima.2011.01.045}}}.

\bibitem[Veloce \em{et~al.}(2016)Veloce, Ku{\'{z}}niak, Stefano, Noble, Boulay,
  Nadeau, Pollmann, Clark, Piquemal, and Schreiner]{Veloce_2016}
Veloce, L.; Ku{\'{z}}niak, M.; Stefano, P.D.; Noble, A.; Boulay, M.; Nadeau,
  P.; Pollmann, T.; Clark, M.; Piquemal, M.; Schreiner, K.
\newblock Temperature dependence of alpha-induced scintillation in the
  1,1,4,4-tetraphenyl-1,3-butadiene wavelength shifter.
\newblock {\em J. Instrum.} {\bf 2016}, {\em 11},~P06003--P06003.
\newblock
  doi:{\changeurlcolor{black}\href{https://doi.org/10.1088/1748-0221/11/06/p06003}{\detokenize{10.1088/1748-0221/11/06/p06003}}}.

\bibitem[Segreto(2015)]{SegretoPhysRevC.91.035503}
Segreto, E.
\newblock Evidence of delayed light emission of tetraphenyl-butadiene excited
  by liquid-argon scintillation light.
\newblock {\em Phys. Rev. C} {\bf 2015}, {\em 91},~035503.
\newblock
  doi:{\changeurlcolor{black}\href{https://doi.org/10.1103/PhysRevC.91.035503}{\detokenize{10.1103/PhysRevC.91.035503}}}.

\bibitem[Segreto \em{et~al.}(2016)Segreto, Machado, Araujo, and
  Teixeira]{Segreto_2016}
Segreto, E.; Machado, A.; Araujo, W.; Teixeira, V.
\newblock Delayed light emission of Tetraphenyl-butadiene excited by liquid
  argon scintillation light. Current status and future plans.
\newblock {\em J. Instrum.} {\bf 2016}, {\em 11},~C02010--C02010.
\newblock
  doi:{\changeurlcolor{black}\href{https://doi.org/10.1088/1748-0221/11/02/c02010}{\detokenize{10.1088/1748-0221/11/02/c02010}}}.

\bibitem[Stanford \em{et~al.}(2018)Stanford, Westerdale, Xu, and
  Calaprice]{PhysRevD.98.062002}
Stanford, C.; Westerdale, S.; Xu, J.; Calaprice, F.
\newblock Surface background suppression in liquid argon dark matter detectors
  using a newly discovered time component of tetraphenyl-butadiene
  scintillation.
\newblock {\em Phys. Rev. D} {\bf 2018}, {\em 98},~062002.
\newblock
  doi:{\changeurlcolor{black}\href{https://doi.org/10.1103/PhysRevD.98.062002}{\detokenize{10.1103/PhysRevD.98.062002}}}.

\bibitem[{Voltz, R.} and {Laustriat, G.}(1968)]{VoltzLaustriat}
{Voltz, R.}.; {Laustriat, G.}.
\newblock Radioluminescence des milieux organiques I. \'Etude cin\'etique.
\newblock {\em J. Phys. France} {\bf 1968}, {\em 29},~159--166.
\newblock
  doi:{\changeurlcolor{black}\href{https://doi.org/10.1051/jphys:01968002902-3015900}{\detokenize{10.1051/jphys:01968002902-3015900}}}.

\bibitem[{Adhikari, P.} and et~al.(2020)]{DeapPulse}
{Adhikari, P.}.; et~al..
\newblock The liquid-argon scintillation pulseshape in DEAP-3600.
\newblock {\em EPJ C} {\bf 2020}, {\em 80},~303.
\newblock
  doi:{\changeurlcolor{black}\href{https://doi.org/10.1140/epjc/s10052-020-7789-x}{\detokenize{10.1140/epjc/s10052-020-7789-x}}}.

\bibitem[Stanford(2017)]{stanford}
Stanford, C.J.
\newblock Alphas and Surface Backgrounds in Liquid Argon Dark Matter Detectors.
\newblock PhD thesis, Princeton University,  2017.
\newblock \url{https://www.proquest.com/docview/2002281264}.

\bibitem[Ouchi \em{et~al.}(2007)Ouchi, Nakai, Ono, and Kimura]{Ouchi}
Ouchi, I.; Nakai, I.; Ono, M.; Kimura, S.
\newblock Features of fluorescence spectra of polyethylene 2,6-naphthalate
  films.
\newblock {\em J. Appl. Polym. Sci.} {\bf 2007}, {\em 105},~114--121.
\newblock
  doi:{\changeurlcolor{black}\href{https://doi.org/10.1002/app.26085}{\detokenize{10.1002/app.26085}}}.

\bibitem[Nakamura \em{et~al.}(2011)Nakamura, Shirakawa, Takahashi, and
  Shimizu]{Nakamura_2011}
Nakamura, H.; Shirakawa, Y.; Takahashi, S.; Shimizu, H.
\newblock Evidence of deep-blue photon emission at high efficiency by common
  plastic.
\newblock {\em EPL} {\bf 2011}, {\em 95},~22001.
\newblock
  doi:{\changeurlcolor{black}\href{https://doi.org/10.1209/0295-5075/95/22001}{\detokenize{10.1209/0295-5075/95/22001}}}.

\bibitem[Efremenko \em{et~al.}(2020)Efremenko, Fajt, Febbraro, Fischer,
  Guitart, Hackett, Hayward, Hod{\'{a}}k, Majorovits, Manzanillas,
  Muenstermann, Öz, Pjatkan, Pohl, Radford, Rouhana, Schulz, {\v{S}}tekl, and
  Stommel]{Efremenko_2020}
Efremenko, Y.; Fajt, L.; Febbraro, M.; Fischer, F.; Guitart, M.; Hackett, B.;
  Hayward, C.; Hod{\'{a}}k, R.; Majorovits, B.; Manzanillas, L.; Muenstermann,
  D.; Öz, E.; Pjatkan, R.; Pohl, M.; Radford, D.; Rouhana, R.; Schulz, O.;
  {\v{S}}tekl, I.; Stommel, M.
\newblock Use of poly(ethylene naphthalate) as a self-vetoing structural
  material.
\newblock {\em J. Phys. Conf. Ser.} {\bf 2020}, {\em 1468},~012225.
\newblock
  doi:{\changeurlcolor{black}\href{https://doi.org/10.1088/1742-6596/1468/1/012225}{\detokenize{10.1088/1742-6596/1468/1/012225}}}.

\bibitem[Soto-Ot\'on(2020)]{SotoOton}
Soto-Ot\'on, J.
\newblock {Measurements of the polyethylene naphthalate performance as a
  wavelength shifter in ProtoDUNE- DP},  2020.
\newblock Presented at Neutrino 2020,
  doi:{\changeurlcolor{black}\href{https://doi.org/10.5281/zenodo.4252545}{\detokenize{10.5281/zenodo.4252545}}}.

\bibitem[{Abt} \em{et~al.}(2020){Abt}, {Dial}, {Efremenko}, {Febbraro},
  {Fischer}, {Guitart}, {Gusev}, {Hackett}, {Hayward}, {Kidder}, {Hodak},
  {Krause}, {Majorovits}, {Manzanillas}, {Muenstermann}, {Pjatkan}, {Pohl},
  {Rouhana}, {Radford}, {Rukhadze}, {Rumyantseva}, {Schilling}, {Schoenert},
  {Schulz}, {Schwarz}, {Stommel}, and {Weingarten}]{penpos}
{Abt}, I.; {Dial}, B.; {Efremenko}, Y.; {Febbraro}, M.; {Fischer}, F.;
  {Guitart}, M.; {Gusev}, K.; {Hackett}, B.; {Hayward}, C.; {Kidder}, M.;
  {Hodak}, R.; {Krause}, P.; {Majorovits}, B.; {Manzanillas}, L.;
  {Muenstermann}, D.; {Pjatkan}, R.; {Pohl}, M.; {Rouhana}, R.; {Radford}, D.;
  {Rukhadze}, E.; {Rumyantseva}, N.; {Schilling}, I.; {Schoenert}, S.;
  {Schulz}, O.; {Schwarz}, M.; {Stommel}, M.; {Weingarten}, J.
\newblock {Usage of PEN as self-vetoing structural material in low background
  experiments}.
\newblock {\em arXiv e-prints} {\bf 2020}, p. arXiv:2011.08983,
  \href{http://arxiv.org/abs/2011.08983}{{\normalfont
  [arXiv:physics.ins-det/2011.08983]}}.

\bibitem[Dorril and et~al(2020)]{ProtoDPspread}
Dorril, R.; et~al.
\newblock Performance studies of PEN and TPB as UV wavelength shifting coatings
  in liquid argon,  2020.
\newblock Presented at Neutrino 2020,
  \url{https://nusoft.fnal.gov/nova/nu2020postersession/pdf/posterPDF-538.pdf},
  doi:{\changeurlcolor{black}\href{https://doi.org/10.5281/zenodo.4123838}{\detokenize{10.5281/zenodo.4123838}}}.

\bibitem[{Janecek}(2012)]{Janecek}
{Janecek}, M.
\newblock Reflectivity Spectra for Commonly Used Reflectors.
\newblock {\em IEEE Trans. Nucl. Sci.} {\bf 2012}, {\em 59},~490--497.
\newblock
  doi:{\changeurlcolor{black}\href{https://doi.org/10.1109/TNS.2012.2183385}{\detokenize{10.1109/TNS.2012.2183385}}}.

\bibitem[Baudis \em{et~al.}(2015)Baudis, Benato, Dressler, Piastra, Usoltsev,
  and Walter]{Baudis_2015}
Baudis, L.; Benato, G.; Dressler, R.; Piastra, F.; Usoltsev, I.; Walter, M.
\newblock Enhancement of light yield and stability of radio-pure
  tetraphenyl-butadiene based coatings for {VUV} light detection in cryogenic
  environments.
\newblock {\em J. Instrum.} {\bf 2015}, {\em 10},~P09009--P09009.
\newblock
  doi:{\changeurlcolor{black}\href{https://doi.org/10.1088/1748-0221/10/09/p09009}{\detokenize{10.1088/1748-0221/10/09/p09009}}}.

\bibitem[{Wang, J.-J}(2017)]{minicleanesr}
{Wang, J.-J}.
\newblock MiniCLEAN Dark Matter Experiment.
\newblock PhD thesis, University of New Mexico,  2017.
\newblock \url{https://digitalrepository.unm.edu/phyc_etds/169}.

\bibitem[Xu(2013)]{tpbtpt}
Xu, J.
\newblock Study of Argon from Underground Sources for Direct Dark Matter
  Detection.
\newblock PhD thesis, Princeton University,  2013.
\newblock \url{http://arks.princeton.edu/ark:/88435/dsp015712m665h}.

\bibitem[Xu \em{et~al.}(2017)Xu, Stanford, Westerdale, Calaprice, Wright, and
  Shi]{tpbtpt0}
Xu, J.; Stanford, C.; Westerdale, S.; Calaprice, F.; Wright, A.; Shi, Z.
\newblock First measurement of surface nuclear recoil background for argon dark
  matter searches.
\newblock {\em Phys. Rev. D} {\bf 2017}, {\em 96},~061101.
\newblock
  doi:{\changeurlcolor{black}\href{https://doi.org/10.1103/PhysRevD.96.061101}{\detokenize{10.1103/PhysRevD.96.061101}}}.

\bibitem[Feng \em{et~al.}(2015)Feng, Jian, Xiang, Yun, Chen, Li, Liu, Hao, and
  Jun]{bismsbchina}
Feng, D.; Jian, W.; Xiang, Z.; Yun, D.; Chen, Y.; Li, Z.; Liu, M.; Hao, C.;
  Jun, C.
\newblock Measurement of the fluorescence quantum yield of bis-MSB.
\newblock {\em Chinese Phys. C} {\bf 2015}, {\em 39}.
\newblock
  doi:{\changeurlcolor{black}\href{https://doi.org/10.1088/1674-1137/39/12/126001}{\detokenize{10.1088/1674-1137/39/12/126001}}}.

\bibitem[Baptista and Mufson(2013)]{BaptistaMSB_2013}
Baptista, B.; Mufson, S.
\newblock Comparison of {TPB} and bis-{MSB} as {VUV} waveshifters in prototype
  {LBNE} photon detector paddles.
\newblock {\em J. Instrum.} {\bf 2013}, {\em 8},~C12003--C12003.
\newblock
  doi:{\changeurlcolor{black}\href{https://doi.org/10.1088/1748-0221/8/12/c12003}{\detokenize{10.1088/1748-0221/8/12/c12003}}}.

\bibitem[Procházka(2016)]{pyreneinps}
Procházka, K.
\newblock {\em Fluorescence Studies of Polymer Containing Systems}; Vol.~16,
  Springer, Cham,  2016.
\newblock
  doi:{\changeurlcolor{black}\href{https://doi.org/10.1007/978-3-319-26788-3}{\detokenize{10.1007/978-3-319-26788-3}}}.

\bibitem[Sahi \em{et~al.}(2018)Sahi, Magill, Ma, Xie, Chen, Jones, and
  Nygren]{nano1}
Sahi, S.; Magill, S.; Ma, L.; Xie, J.; Chen, W.; Jones, B.; Nygren, D.
\newblock Wavelength-shifting properties of luminescence nanoparticles for high
  energy particle detection and specific physics process observation.
\newblock {\em Sci. Rep.} {\bf 2018}, {\em 8},~10515.
\newblock
  doi:{\changeurlcolor{black}\href{https://doi.org/10.1038/s41598-018-28741-y}{\detokenize{10.1038/s41598-018-28741-y}}}.

\bibitem[Datta \em{et~al.}(2020)Datta, Barman, Magill, and Motakef]{perovskite}
Datta, A.; Barman, B.; Magill, S.; Motakef, S.
\newblock Highly efficient photon detection systems for noble liquid detectors
  based on perovskite quantum dots.
\newblock {\em Sci. Rep.} {\bf 2020}, {\em 10},~16932.
\newblock
  doi:{\changeurlcolor{black}\href{https://doi.org/10.1038/s41598-020-73437-x}{\detokenize{10.1038/s41598-020-73437-x}}}.

\bibitem[Peiffer \em{et~al.}(2008)Peiffer, Pollmann, Schönert, Smolnikov, and
  Vasiliev]{Peiffer_2008}
Peiffer, P.; Pollmann, T.; Schönert, S.; Smolnikov, A.; Vasiliev, S.
\newblock Pulse shape analysis of scintillation signals from pure and
  xenon-doped liquid argon for radioactive background identification.
\newblock {\em J. Instrum.} {\bf 2008}, {\em 3},~P08007--P08007.
\newblock
  doi:{\changeurlcolor{black}\href{https://doi.org/10.1088/1748-0221/3/08/p08007}{\detokenize{10.1088/1748-0221/3/08/p08007}}}.

\bibitem[Wahl \em{et~al.}(2014)Wahl, Bernard, Lippincott, Nikkel, Shin, and
  McKinsey]{Lippincott_2014}
Wahl, C.G.; Bernard, E.P.; Lippincott, W.H.; Nikkel, J.A.; Shin, Y.; McKinsey,
  D.N.
\newblock Pulse-shape discrimination and energy resolution of a liquid-argon
  scintillator with xenon doping.
\newblock {\em J. Instrum.} {\bf 2014}, {\em 9},~P06013--P06013.
\newblock
  doi:{\changeurlcolor{black}\href{https://doi.org/10.1088/1748-0221/9/06/p06013}{\detokenize{10.1088/1748-0221/9/06/p06013}}}.

\bibitem[Akimov \em{et~al.}(2019)Akimov, Belov, Konovalov, Kumpan, Razuvaeva,
  Rudik, and Simakov]{Rudik_2019}
Akimov, D.; Belov, V.; Konovalov, A.; Kumpan, A.; Razuvaeva, O.; Rudik, D.;
  Simakov, G.
\newblock Fast component re-emission in Xe-doped liquid argon.
\newblock {\em J. Instrum.} {\bf 2019}, {\em 14},~P09022--P09022.
\newblock
  doi:{\changeurlcolor{black}\href{https://doi.org/10.1088/1748-0221/14/09/p09022}{\detokenize{10.1088/1748-0221/14/09/p09022}}}.

\bibitem[Galbiati \em{et~al.}(2020)Galbiati, Li, Luo, Marlow, Wang, and
  Wang]{Galbiati:2020eup}
Galbiati, C.; Li, X.; Luo, J.; Marlow, D.R.; Wang, H.; Wang, Y.
\newblock {Pulse shape study of the fast scintillation light emitted from
  xenon-doped liquid argon using silicon photomultipliers}.
\newblock {\em arXiv e-prints} {\bf 2020},
  \href{http://arxiv.org/abs/2009.06238}{{\normalfont
  [arXiv:physics.ins-det/2009.06238]}}.

\bibitem[{Amaudruz, P.-A} and et~al(2015)]{deap1}
{Amaudruz, P.-A}.; et~al.
\newblock Radon backgrounds in the DEAP-1 liquid-argon-based Dark Matter
  detector.
\newblock {\em Astropart. Phys.} {\bf 2015}, {\em 62},~178 -- 194.
\newblock
  doi:{\changeurlcolor{black}\href{https://doi.org/10.1016/j.astropartphys.2014.09.006}{\detokenize{10.1016/j.astropartphys.2014.09.006}}}.

\bibitem[Aalseth and et~al(2020)]{dart}
Aalseth, C.; et~al.
\newblock Design and construction of a new detector to measure ultra-low
  radioactive-isotope contamination of argon.
\newblock {\em J. Instrum.} {\bf 2020}, {\em 15},~P02024--P02024.
\newblock
  doi:{\changeurlcolor{black}\href{https://doi.org/10.1088/1748-0221/15/02/p02024}{\detokenize{10.1088/1748-0221/15/02/p02024}}}.

\bibitem[Cs\'athy \em{et~al.}(2016)Cs\'athy, Bode, Kratz, Sch\"onert, and
  Wiesinger]{Csathy:2016wdy}
Cs\'athy, J.J.; Bode, T.; Kratz, J.; Sch\"onert, S.; Wiesinger, C.
\newblock {Optical fiber read-out for liquid argon scintillation light}.
\newblock {\em arXiv e-prints} {\bf 2016},
  \href{http://arxiv.org/abs/1606.04254}{{\normalfont
  [arXiv:physics.ins-det/1606.04254]}}.

\bibitem[Boccone and et~al(2009)]{ArDM_refl_2009}
Boccone, V.; et~al.
\newblock Development of wavelength shifter coated reflectors for the {ArDM}
  argon dark matter detector.
\newblock {\em J. Instrum.} {\bf 2009}, {\em 4},~P06001--P06001.
\newblock
  doi:{\changeurlcolor{black}\href{https://doi.org/10.1088/1748-0221/4/06/p06001}{\detokenize{10.1088/1748-0221/4/06/p06001}}}.

\bibitem[Yang \em{et~al.}(2020)Yang, Xu, Tang, and Zhang]{Yang:2019zob}
Yang, H.; Xu, Z.F.; Tang, J.; Zhang, Y.
\newblock {Spin coating of TPB film on acrylic substrate and measurement of its
  wavelength shifting efficiency}.
\newblock {\em Nucl. Sci. Tech.} {\bf 2020}, {\em 31},~28,
  \href{http://arxiv.org/abs/1911.08897}{{\normalfont
  [arXiv:physics.ins-det/1911.08897]}}.
\newblock
  doi:{\changeurlcolor{black}\href{https://doi.org/10.1007/s41365-020-0737-5}{\detokenize{10.1007/s41365-020-0737-5}}}.

\bibitem[Bower(2002)]{Bower:994867}
Bower, K.E.
\newblock {\em {Polymers, Phosphors, and Voltaics for Radioisotope
  Microbatteries}}; CRC Press: Abingdon,  2002.

\bibitem[Mavrokoridis(2011)]{Mavrokoridis_2011}
Mavrokoridis, K.
\newblock Light Readout Optimisation using Wavelength Shifter - Reflector
  Combinations.
\newblock {\em J. Phys. Conf. Ser.} {\bf 2011}, {\em 308},~012020.
\newblock
  doi:{\changeurlcolor{black}\href{https://doi.org/10.1088/1742-6596/308/1/012020}{\detokenize{10.1088/1742-6596/308/1/012020}}}.

\bibitem[Baptista \em{et~al.}(2014)Baptista, Bugel, Chiu, Conrad, Ignarra,
  Jones, Katori, and Mufson]{baptista2014benchmarking}
Baptista, B.; Bugel, L.; Chiu, C.; Conrad, J.M.; Ignarra, C.M.; Jones, B.J.P.;
  Katori, T.; Mufson, S.
\newblock Benchmarking TPB-coated Light Guides for Liquid Argon TPC Light
  Detection Systems,  2014,  \href{http://arxiv.org/abs/1210.3793}{{\normalfont
  [arXiv:physics.ins-det/1210.3793]}}.

\bibitem[Moss \em{et~al.}(2016)Moss, Moon, Bugel, Conrad, Sachdev, Toups, and
  Wongjirad]{moss2016factor}
Moss, Z.; Moon, J.; Bugel, L.; Conrad, J.M.; Sachdev, K.; Toups, M.; Wongjirad,
  T.
\newblock A Factor of Four Increase in Attenuation Length of Dipped Lightguides
  for Liquid Argon TPCs Through Improved Coating,  2016,
  \href{http://arxiv.org/abs/1604.03103}{{\normalfont
  [arXiv:physics.ins-det/1604.03103]}}.

\bibitem[{Lubashevskiy, A} \em{et~al.}(2018){Lubashevskiy, A}, {Agostini, M},
  {Budj\'as, D}, {Gangapshev, A}, {Gusev, K}, {Heisel, M}, {Klimenko, A},
  {Lazzaro, A}, {Lehnert, B}, {Pelczar, K}, {Sch\"onert, S}, {Smolnikov, A},
  {Walter, M}, and {Zuzel, G}]{GerdaEnclosures}
{Lubashevskiy, A}.; {Agostini, M}.; {Budj\'as, D}.; {Gangapshev, A}.; {Gusev,
  K}.; {Heisel, M}.; {Klimenko, A}.; {Lazzaro, A}.; {Lehnert, B}.; {Pelczar,
  K}.; {Sch\"onert, S}.; {Smolnikov, A}.; {Walter, M}.; {Zuzel, G}.
\newblock Mitigation of $^{42}$Ar/$^{42}$K background for the GERDA Phase II
  experiment.
\newblock {\em EPJ C} {\bf 2018}, {\em 78},~15.
\newblock
  doi:{\changeurlcolor{black}\href{https://doi.org/10.1140/epjc/s10052-017-5499-9}{\detokenize{10.1140/epjc/s10052-017-5499-9}}}.

\bibitem[Sanguino \em{et~al.}(2016)Sanguino, Balau, {Botelho do Rego}, Pereira,
  and Chepel]{xetpb}
Sanguino, P.; Balau, F.; {Botelho do Rego}, A.; Pereira, A.; Chepel, V.
\newblock Stability of tetraphenyl butadiene thin films in liquid xenon.
\newblock {\em Thin Solid Films} {\bf 2016}, {\em 600},~65 -- 70.
\newblock
  doi:{\changeurlcolor{black}\href{https://doi.org/10.1016/j.tsf.2016.01.006}{\detokenize{10.1016/j.tsf.2016.01.006}}}.

\bibitem[Asaadi \em{et~al.}(2019)Asaadi, Jones, Tripathi, Parmaksiz, Sullivan,
  and Williams]{tpbdissolved}
Asaadi, J.; Jones, B.; Tripathi, A.; Parmaksiz, I.; Sullivan, H.; Williams, Z.
\newblock Emanation and bulk fluorescence in liquid argon from tetraphenyl
  butadiene wavelength shifting coatings.
\newblock {\em J. Instrum.} {\bf 2019}, {\em 14},~P02021--P02021.
\newblock
  doi:{\changeurlcolor{black}\href{https://doi.org/10.1088/1748-0221/14/02/p02021}{\detokenize{10.1088/1748-0221/14/02/p02021}}}.

\bibitem[Acciarri \em{et~al.}(2013)Acciarri, Canci, Cavanna, Segreto, and
  Szelc]{Acciarri_2013}
Acciarri, R.; Canci, N.; Cavanna, F.; Segreto, E.; Szelc, A.M.
\newblock Aging studies on thin tetra-phenyl butadiene films.
\newblock {\em J. Instrum.} {\bf 2013}, {\em 8},~C10002--C10002.
\newblock
  doi:{\changeurlcolor{black}\href{https://doi.org/10.1088/1748-0221/8/10/c10002}{\detokenize{10.1088/1748-0221/8/10/c10002}}}.

\bibitem[Chiu \em{et~al.}(2012)Chiu, Ignarra, Bugel, Chen, Conrad, Jones,
  Katori, and Moult]{Chiu_2012}
Chiu, C.S.; Ignarra, C.; Bugel, L.; Chen, H.; Conrad, J.M.; Jones, B.J.P.;
  Katori, T.; Moult, I.
\newblock Environmental effects on {TPB} wavelength-shifting coatings.
\newblock {\em J. Instrum.} {\bf 2012}, {\em 7},~P07007--P07007.
\newblock
  doi:{\changeurlcolor{black}\href{https://doi.org/10.1088/1748-0221/7/07/p07007}{\detokenize{10.1088/1748-0221/7/07/p07007}}}.

\bibitem[Yahlali \em{et~al.}(2017)Yahlali, Garcia, Díaz, Soriano, and
  Fernandes]{YAHLALI2017109}
Yahlali, N.; Garcia, J.; Díaz, J.; Soriano, A.; Fernandes, L.
\newblock Ageing studies of TPB in noble gas detectors for dark matter and
  neutrinoless $\beta\beta$ decay searches.
\newblock {\em Spectrochim. Acta A} {\bf 2017}, {\em 172},~109 -- 114.
\newblock Special Issue: Colloquium Spectroscopicum Internationale XXXIX,
  doi:{\changeurlcolor{black}\href{https://doi.org/10.1016/j.saa.2016.04.025}{\detokenize{10.1016/j.saa.2016.04.025}}}.

\bibitem[Jones \em{et~al.}(2013)Jones, VanGemert, Conrad, and
  Pla-Dalmau]{Jones_2013}
Jones, B.J.P.; VanGemert, J.K.; Conrad, J.M.; Pla-Dalmau, A.
\newblock Photodegradation mechanisms of tetraphenyl butadiene coatings for
  liquid argon detectors.
\newblock {\em J. Instrum.} {\bf 2013}, {\em 8},~P01013--P01013.
\newblock
  doi:{\changeurlcolor{black}\href{https://doi.org/10.1088/1748-0221/8/01/p01013}{\detokenize{10.1088/1748-0221/8/01/p01013}}}.

\bibitem[Macdonald \em{et~al.}(2007)Macdonald, Looney, MacKerron, Eveson, Adam,
  Hashimoto, and Rakos]{penoligomers}
Macdonald, W.; Looney, M.; MacKerron, D.; Eveson, R.; Adam, R.; Hashimoto, K.;
  Rakos, K.
\newblock Latest Advances in Substrates for Flexible Electronics.
\newblock {\em J. Soc. Inf. Disp.} {\bf 2007}, {\em 15},~1075--1083.
\newblock
  doi:{\changeurlcolor{black}\href{https://doi.org/10.1889/1.2825093}{\detokenize{10.1889/1.2825093}}}.

\bibitem[{Mary} \em{et~al.}(2001){Mary}, {Teyssedre}, and {Laurent}]{penaging}
{Mary}, D.; {Teyssedre}, G.; {Laurent}, C.
\newblock UV-induced degradation of poly(ethylene naphthalate) films from the
  standpoint of electrical and luminescence properties.
\newblock  2001 Annual Report Conference on Electrical Insulation and
  Dielectric Phenomena (Cat. No.01CH37225),  2001, pp. 165--168.
\newblock
  doi:{\changeurlcolor{black}\href{https://doi.org/10.1109/CEIDP.2001.963512}{\detokenize{10.1109/CEIDP.2001.963512}}}.

\bibitem[{Bolozdynya} \em{et~al.}(2008){Bolozdynya}, {Bradley}, {Brusov},
  {Dahl}, {Kwong}, and {Shutt}]{ptpsolubility}
{Bolozdynya}, A.I.; {Bradley}, A.W.; {Brusov}, P.P.; {Dahl}, C.E.; {Kwong}, J.;
  {Shutt}, T.
\newblock Using a Wavelength Shifter to Enhance the Sensitivity of Liquid Xenon
  Dark Matter Detectors.
\newblock {\em IEEE Trans. Nucl. Sci.} {\bf 2008}, {\em 55},~1453--1457.
\newblock
  doi:{\changeurlcolor{black}\href{https://doi.org/10.1109/TNS.2008.919258}{\detokenize{10.1109/TNS.2008.919258}}}.

\bibitem[Garcia-Gamez \em{et~al.}(2020)Garcia-Gamez, Green, and
  Szelc]{Garcia-Gamez:2020xrv}
Garcia-Gamez, D.; Green, P.; Szelc, A.
\newblock {Predicting Transport Effects of Scintillation Light Signals in
  Large-Scale Liquid Argon Detectors}.
\newblock {\em arXiv e-prints} {\bf 2020},
  \href{http://arxiv.org/abs/2010.00324}{{\normalfont
  [arXiv:physics.ins-det/2010.00324]}}.

\bibitem[Sorel(2014)]{Sorel_2014}
Sorel, M.
\newblock Expected performance of an ideal liquid argon neutrino detector with
  enhanced sensitivity to scintillation light.
\newblock {\em J. Instrum.} {\bf 2014}, {\em 9},~P10002--P10002.
\newblock
  doi:{\changeurlcolor{black}\href{https://doi.org/10.1088/1748-0221/9/10/p10002}{\detokenize{10.1088/1748-0221/9/10/p10002}}}.

\bibitem[Agnes and et~al(2015)]{ds-50}
Agnes, P.; et~al.
\newblock First results from the DarkSide-50 dark matter experiment at
  Laboratori Nazionali del Gran Sasso.
\newblock {\em Phys. Lett. B} {\bf 2015}, {\em 743},~456 -- 466.
\newblock
  doi:{\changeurlcolor{black}\href{https://doi.org/10.1016/j.physletb.2015.03.012}{\detokenize{10.1016/j.physletb.2015.03.012}}}.

\bibitem[Machado \em{et~al.}(2019)Machado, Palamara, and
  Schmitz]{Machado:2019oxb}
Machado, P.A.; Palamara, O.; Schmitz, D.W.
\newblock {The Short-Baseline Neutrino Program at Fermilab}.
\newblock {\em Ann. Rev. Nucl. Part. Sci.} {\bf 2019}, {\em 69},~363--387,
  \href{http://arxiv.org/abs/1903.04608}{{\normalfont
  [arXiv:hep-ex/1903.04608]}}.
\newblock
  doi:{\changeurlcolor{black}\href{https://doi.org/10.1146/annurev-nucl-101917-020949}{\detokenize{10.1146/annurev-nucl-101917-020949}}}.

\bibitem[Szelc \em{et~al.}(2020)Szelc, Basque, Garcia-Gamez, Segreto,
  Littlejohn, Dorrill, Foreman, Cavanna, and Ku\'zniak]{wlsLOI}
Szelc, A.M.; Basque, V.; Garcia-Gamez, D.; Segreto, E.; Littlejohn, B.;
  Dorrill, R.; Foreman, W.; Cavanna, F.; Ku\'zniak, M.
\newblock Wavelength-shifting Reflector Foils in Liquid Argon Neutrino
  Detectors,  2020.
\newblock Snowmass 2021 LOI,
  \url{https://www.snowmass21.org/docs/files/summaries/IF/SNOWMASS21-IF8_IF2_Andrzej_Szelc-145.pdf}.

\bibitem[Spanu \em{et~al.}(2018)Spanu, Falcone, Mazza, Menegolli, Prata,
  Raselli, Rossella, and Torti]{icarustpb}
Spanu, M.; Falcone, A.; Mazza, R.; Menegolli, A.; Prata, M.C.; Raselli, G.L.;
  Rossella, M.; Torti, M.
\newblock Study on {TPB} as wavelength shifter for the new {ICARUS} T600 light
  collection system in the {SBN} program.
\newblock {\em J. Phys. Conf. Ser.} {\bf 2018}, {\em 956},~012016.
\newblock
  doi:{\changeurlcolor{black}\href{https://doi.org/10.1088/1742-6596/956/1/012016}{\detokenize{10.1088/1742-6596/956/1/012016}}}.

\bibitem[Bugel \em{et~al.}(2011)Bugel, Conrad, Ignarra, Jones, Katori, Smidt,
  and Tanaka]{Bugel_2011}
Bugel, L.; Conrad, J.; Ignarra, C.; Jones, B.; Katori, T.; Smidt, T.; Tanaka,
  H.K.
\newblock Demonstration of a lightguide detector for liquid argon TPCs.
\newblock {\em Nucl. Instrum. Methods Phys. Res. A} {\bf 2011}, {\em
  640},~69–75.
\newblock
  doi:{\changeurlcolor{black}\href{https://doi.org/10.1016/j.nima.2011.03.003}{\detokenize{10.1016/j.nima.2011.03.003}}}.

\bibitem[Abi and et~al(2020)]{ProtoDUNErun}
Abi, B.; et~al.
\newblock {First results on ProtoDUNE-SP liquid argon time projection chamber
  performance from a beam test at the CERN Neutrino Platform}.
\newblock {\em arXiv e-prints} {\bf 2020},
  \href{http://arxiv.org/abs/2007.06722}{{\normalfont
  [arXiv:physics.ins-det/2007.06722]}}.

\bibitem[Howard \em{et~al.}(2018)Howard, Mufson, Whittington, Adams, Baugh,
  Jordan, Karty, Macias, and Pla-Dalmau]{HOWARD20189}
Howard, B.; Mufson, S.; Whittington, D.; Adams, B.; Baugh, B.; Jordan, J.;
  Karty, J.; Macias, C.; Pla-Dalmau, A.
\newblock A novel use of light guides and wavelength shifting plates for the
  detection of scintillation photons in large liquid argon detectors.
\newblock {\em Nucl. Instrum. Methods Phys. Res. A} {\bf 2018}, {\em 907},~9 --
  21.
\newblock Advances in Instrumentation and Experimental Methods (Special Issue
  in Honour of Kai Siegbahn),
  doi:{\changeurlcolor{black}\href{https://doi.org/10.1016/j.nima.2018.06.050}{\detokenize{10.1016/j.nima.2018.06.050}}}.

\bibitem[Howard(2018)]{Howard_2018}
Howard, B.
\newblock Liquid argon scintillation detection utilizing wavelength-shifting
  plates and light guides.
\newblock {\em J. Instrum.} {\bf 2018}, {\em 13},~C02006--C02006.
\newblock
  doi:{\changeurlcolor{black}\href{https://doi.org/10.1088/1748-0221/13/02/c02006}{\detokenize{10.1088/1748-0221/13/02/c02006}}}.

\bibitem[Bazetto \em{et~al.}(2020)Bazetto, Pimentel, Machado, and
  Segreto]{arapuca}
Bazetto, M.Q.; Pimentel, V.; Machado, A.; Segreto, E.
\newblock Study and characterization of {WLS} for {ARAPUCA} to {DUNE}
  experiment.
\newblock {\em J. Instrum.} {\bf 2020}, {\em 15},~C05048--C05048.
\newblock
  doi:{\changeurlcolor{black}\href{https://doi.org/10.1088/1748-0221/15/05/c05048}{\detokenize{10.1088/1748-0221/15/05/c05048}}}.

\bibitem[Segreto \em{et~al.}(2020)Segreto, Machado, Fauth, Ramos, de~Souza,
  Souza, Pimentel, Bazetto, and Ayala-Torres]{xarapuca}
Segreto, E.; Machado, A.; Fauth, A.; Ramos, R.; de~Souza, G.; Souza, H.;
  Pimentel, V.; Bazetto, M.Q.; Ayala-Torres, M.
\newblock First liquid argon test of the X-{ARAPUCA}.
\newblock {\em J. Instrum.} {\bf 2020}, {\em 15},~C05045--C05045.
\newblock
  doi:{\changeurlcolor{black}\href{https://doi.org/10.1088/1748-0221/15/05/c05045}{\detokenize{10.1088/1748-0221/15/05/c05045}}}.

\bibitem[Auger \em{et~al.}(2018)Auger, Chen, Ereditato, Goeldi, Kreslo, Lorca,
  Luethi, Mettler, Sinclair, and Weber]{arclight}
Auger, M.; Chen, Y.; Ereditato, A.; Goeldi, D.; Kreslo, I.; Lorca, D.; Luethi,
  M.; Mettler, T.; Sinclair, J.; Weber, M.
\newblock {ArCLight\textemdash{}A Compact Dielectric Large-Area Photon
  Detector}.
\newblock {\em Instruments} {\bf 2018}, {\em 2},~3,
  \href{http://arxiv.org/abs/1711.11409}{{\normalfont
  [arXiv:physics.ins-det/1711.11409]}}.
\newblock
  doi:{\changeurlcolor{black}\href{https://doi.org/10.3390/instruments2010003}{\detokenize{10.3390/instruments2010003}}}.

\bibitem[Auger \em{et~al.}(2019)Auger, Berner, Chen, Ereditato, Goeldi, Koller,
  Kreslo, Lorca, Mettler, Piastra, Sinclair, Weber, Wilkinson, Convery, Domine,
  Drielsma, Itay, Koh, Tanaka, Terao, Tsang, Usher, Dwyer, Kohn, Madigan,
  Marshall, Bross, and Asaadi]{auger2019new}
Auger, M.; Berner, R.; Chen, Y.; Ereditato, A.; Goeldi, D.; Koller, P.P.;
  Kreslo, I.; Lorca, D.; Mettler, T.; Piastra, F.; Sinclair, J.R.; Weber, M.;
  Wilkinson, C.; Convery, M.; Domine, L.; Drielsma, F.; Itay, R.; Koh, D.H.;
  Tanaka, H.A.; Terao, K.; Tsang, P.; Usher, T.; Dwyer, D.A.; Kohn, S.;
  Madigan, P.; Marshall, C.M.; Bross, A.; Asaadi, J.
\newblock A New Concept for Kilotonne Scale Liquid Argon Time Projection
  Chambers,  2019,  \href{http://arxiv.org/abs/1908.10956}{{\normalfont
  [arXiv:physics.ins-det/1908.10956]}}.

\bibitem[Boulay and Hime(2006)]{Boulay:2006mb}
Boulay, M.; Hime, A.
\newblock {Technique for direct detection of weakly interacting massive
  particles using scintillation time discrimination in liquid argon}.
\newblock {\em Astropart. Phys.} {\bf 2006}, {\em 25},~179--182.
\newblock
  doi:{\changeurlcolor{black}\href{https://doi.org/10.1016/j.astropartphys.2005.12.009}{\detokenize{10.1016/j.astropartphys.2005.12.009}}}.

\bibitem[Hitachi \em{et~al.}(1983)Hitachi, Takahashi, Funayama, Masuda,
  Kikuchi, and Doke]{hitachi}
Hitachi, A.; Takahashi, T.; Funayama, N.; Masuda, K.; Kikuchi, J.; Doke, T.
\newblock Effect of ionization density on the time dependence of luminescence
  from liquid argon and xenon.
\newblock {\em Phys. Rev. B} {\bf 1983}, {\em 27},~5279--5285.
\newblock
  doi:{\changeurlcolor{black}\href{https://doi.org/10.1103/PhysRevB.27.5279}{\detokenize{10.1103/PhysRevB.27.5279}}}.

\bibitem[{Amaudruz, P.-A} and et~al(2016)]{AMAUDRUZ20161}
{Amaudruz, P.-A}.; et~al.
\newblock Measurement of the scintillation time spectra and pulse-shape
  discrimination of low-energy $\beta$ and nuclear recoils in liquid argon with
  DEAP-1.
\newblock {\em Astropart. Phys.} {\bf 2016}, {\em 85},~1 -- 23.
\newblock
  doi:{\changeurlcolor{black}\href{https://doi.org/10.1016/j.astropartphys.2016.09.002}{\detokenize{10.1016/j.astropartphys.2016.09.002}}}.

\bibitem[Acciarri and et~al(2010)]{Acciarri_2010}
Acciarri, R.; et~al.
\newblock Effects of Nitrogen contamination in liquid Argon.
\newblock {\em J. Instrum.} {\bf 2010}, {\em 5},~P06003--P06003.
\newblock
  doi:{\changeurlcolor{black}\href{https://doi.org/10.1088/1748-0221/5/06/p06003}{\detokenize{10.1088/1748-0221/5/06/p06003}}}.

\bibitem[Agnes and et~al(2018)]{PhysRevD.98.102006}
Agnes, P.; et~al.
\newblock DarkSide-50 532-day dark matter search with low-radioactivity argon.
\newblock {\em Phys. Rev. D} {\bf 2018}, {\em 98},~102006.
\newblock
  doi:{\changeurlcolor{black}\href{https://doi.org/10.1103/PhysRevD.98.102006}{\detokenize{10.1103/PhysRevD.98.102006}}}.

\bibitem[Wang(2014)]{boqianPhD}
Wang, B.
\newblock Alpha Background Study In Dark Matter Detection.
\newblock PhD thesis, Syracuse University,  2014.
\newblock \url{https://surface.syr.edu/etd/194}.

\bibitem[Boulay and Ku\'zniak(2020)]{surfacepsd}
Boulay, M.; Ku\'zniak, M.
\newblock Technique for surface background rejection in liquid argon dark
  matter detectors using layered wavelength-shifting and scintillating thin
  films.
\newblock {\em Nucl. Instrum. Methods Phys. Res. A} {\bf 2020}, {\em
  968},~163631.
\newblock
  doi:{\changeurlcolor{black}\href{https://doi.org/10.1016/j.nima.2020.163631}{\detokenize{10.1016/j.nima.2020.163631}}}.

\bibitem[Gallacher and Boulay(2020)]{surfacepsd2}
Gallacher, D.; Boulay, M.
\newblock Surface background rejection technique for liquid argon dark matter
  detectors using a thin scintillating layer.
\newblock {\em J. Instrum.} {\bf 2020}, {\em 15},~C03016--C03016.
\newblock
  doi:{\changeurlcolor{black}\href{https://doi.org/10.1088/1748-0221/15/03/c03016}{\detokenize{10.1088/1748-0221/15/03/c03016}}}.

\bibitem[Clark \em{et~al.}(2016)Clark, Ku\'zniak, Zheng, and {Di
  Stefano}]{captalk}
Clark, M.; Ku\'zniak, M.; Zheng, M.; {Di Stefano}, P.
\newblock {Spectroscopic and time-resolved measurements of the fluorescence of
  pyrene at low temperatures for noble liquid particle detectors},  2016.
\newblock Presented at CAP 2016,
  \url{https://indico.cern.ch/event/472838/contributions/1150228}.

\bibitem[Ajaj and et~al(2019)]{deap231}
Ajaj, R.; et~al.
\newblock Search for dark matter with a 231-day exposure of liquid argon using
  DEAP-3600 at SNOLAB.
\newblock {\em Phys. Rev. D} {\bf 2019}, {\em 100},~022004.
\newblock
  doi:{\changeurlcolor{black}\href{https://doi.org/10.1103/PhysRevD.100.022004}{\detokenize{10.1103/PhysRevD.100.022004}}}.

\bibitem[doe(2020)]{doe}
DOE Basic Research Needs Study on High Energy Physics Detector Research and
  Development,  2020.
\newblock
  \url{https://science.osti.gov/-/media/hep/pdf/Reports/2020/DOE_Basic_Research_Needs_Study_on_High_Energy_Physics.pdf}.

\end{thebibliography}

\end{document}